\documentclass[apj]{emulateapj}
\usepackage{xspace,graphics,graphicx}
\usepackage{lscape}

\newcommand\msun{\ensuremath{M_\sun}\xspace}

\newcommand\teff{\ensuremath{T_{\rm eff}}\xspace}
\newcommand\logg{\ensuremath{\log g}\xspace}
\newcommand\ubv{\ensuremath{U\!BV}}
\newcommand\ebv{\ensuremath{E(\bv)}\xspace}

\journalinfo{}
\slugcomment{Accepted for publication in the Astronomical Journal 2008 March 28} 
\shorttitle{White dwarfs in NGC~1039 (M34)}
\shortauthors{Rubin et al.}

\begin{document}

\title{The white dwarf population in NGC~1039 (M34) and the white dwarf initial-final mass relation\footnotemark[*]}
\footnotetext[*]{Some of the data presented herein were obtained at the W. M. Keck Observatory, which is operated as a scientific partnership among the California Institute of Technology, the University of California and the National Aeronautics and Space Administration.  The Observatory was made possible by the generous financial support of the W.M. Keck Foundation.}

\author{Kate H. R. Rubin\altaffilmark{1}, Kurtis A. Williams\altaffilmark{2}, M. Bolte\altaffilmark{1}, \& Detlev Koester\altaffilmark{3}}

\altaffiltext{1}{University of California Observatories, University of California, Santa Cruz, CA 95064; rubin@ucolick.org, bolte@ucolick.org}
\altaffiltext{2}{NSF Astronomy \& Astrophysics Postdoctoral Fellow and Department of Astronomy, University of Texas, Austin, TX 78712; kurtis@astro.as.utexas.edu}
\altaffiltext{3}{Institut f\"ur Theoretische Physik und Astrophysik, Universit\"at Kiel, 24098 Kiel, Germany; 
koester@astrophysik.uni-kiel.de}

\begin{abstract}
  We present the first detailed photometric and spectroscopic study of
  the white dwarfs (WDs) in the field of the $\sim 225$ Myr old
  ($\log \tau_{cl} = 8.35$) 
  open cluster NGC 1039 (M34) as part of the
  ongoing Lick-Arizona White Dwarf Survey.  Using wide-field $\ubv$
  imaging, we photometrically select 44 WD candidates in this field.
  We spectroscopically identify 19 of these objects as WDs; 17
  are hydrogen-atmosphere DA WDs, one is a helium-atmosphere DB WD,
  and one is a cool DC WD that exhibits no detectable absorption
  lines.  
  We find an effective temperature ($\teff$) and surface gravity
  ($\logg$) for each DA WD by fitting Balmer-line profiles
  from model atmospheres to the observed spectra.  WD evolutionary
  models are then invoked to derive masses and cooling times for each
  DA WD.  Of the 17 DAs,
  five are at the approximate distance modulus of the cluster.  Another
  WD with a distance modulus $0.45$ mag brighter than that of the
  cluster could be a double-degenerate binary cluster
  member, but is more likely to be a field WD.  
  We place the five single cluster member WDs in the empirical 
  initial-final mass relation and find that
  three of them lie very close to the previously derived linear
  relation; two have WD masses significantly below the relation.
  These outliers may have experienced some sort of enhanced mass loss
  or binary evolution; however, it is quite possible that these WDs are
  simply interlopers from the field WD population.
Eight of the 17 DA WDs show significant \ion{Ca}{2} K absorption; 
comparison of the absorption strength with the WD distances suggests
  that the absorption is interstellar, though this cannot be confirmed
  with the current data.

\end{abstract}

\section{Introduction}
This is the fourth installment in a series of papers investigating the
white dwarf (WD) content in young open clusters.  The primary goal of
these studies is to better constrain the integrated mass lost by low
to intermediate mass stars during their main sequence and advanced
stages of evolution.  This mass loss is most commonly represented as
an initial-final mass relation \citep[IFMR,
e.g.,][]{Weidemann1977,Weidemann2000}.  Constraining the IFMR is
crucial to understanding the chemical enrichment due to stellar mass
loss and star formation efficiency in galaxies \citep{Ferrario2005}.
The shape of the IFMR has a direct effect on the shape of the WD
luminosity function (WDLF), which is commonly used to constrain the
age of the Galactic thin disk \citep[e.g.,][]{Winget1987,Harris2006},
the thick disk, the Galactic halo \citep[e.g.][]{vonHippel2005b}, and
globular clusters \citep[e.g.,][]{Hansen2002,Hansen2004,Hansen2007}.

Much work has been done to constrain the IFMR semi-empirically.  The
first purely photographic study was done by \citet{Romanishin1980};
spectroscopy of WDs, which is necessary to determine their masses and
cooling times, was added to studies of the IFMR in a series of papers by
\citet[e.g.,][]{Koester1981,Koester1996}.  Motivated by this progress,
\citet{Weidemann2000} updated his semi-empirical IFMR.  More recent
work on open clusters by \citet{Claver2001}, \citet{Kalirai2005} and
\citet{Williams2004a} led \citet{Ferrario2005} to re-determine the
empirical IFMR.  They found that a linear function adequately
described the observational data.  More recent purely empirical IFMR
determinations \citep[e.g.,][]{Williams2007c,Kalirai2008} also find
the empirical IFMR can be adequately parameterized as a linear
function; the slopes derived in these studies are consistent with the slope
found by \citet{Ferrario2005}.

Another aim of these studies is to more accurately determine the upper
limit on the main sequence mass of a star that evolves into a WD.
This limit is an important parameter in modeling dwarf galaxy
evolution in particular \citep{Dekel1986}, as it constrains the amount
of energy and matter produced during the evolution of a given stellar
population.  Recent theoretical studies \citep{Poelarends2007} of the
advanced stages of stellar evolution find that the lower limit on the
mass of a solar metallicity star that will ignite carbon burning is
$\sim 7 M_{\odot}$.  Only stars with masses below this limit should 
be able to form CO-core WDs.  Between $\sim 7 M_{\odot}$ and $\sim 9
M_{\odot}$, stars modeled in \citet{Poelarends2007} ignite carbon but
do not ignite neon and oxygen, so that some of them may form ONe-core
WDs.  These limits have yet to be tested observationally.  WD studies
to date have investigated WDs with initial masses between $\sim
1.2$ and $6.5 M_{\odot}$ \citep{Ferrario2005, Kalirai2008}.  It is important to explore
the WD populations of \it{young} \rm open clusters with possible
massive, still relatively bright WDs for this reason.

A third use of these studies of WDs in young open clusters is to
compare cluster ages as determined by main-sequence fitting with WD
cooling times \citep{vonHippel2005a}.  Consistency of these ages
increases confidence in both stellar evolutionary models and WD
cooling models.  Consistency across a wide range of star cluster ages
further increases confidence in studies seeking to derive star-formation 
history from the WDLF \citep[e.g.,][]{Liebert2005}.

Our general approach is to use photometric observations of young
clusters in order to select likely WD cluster member candidates based
on their location in the color-magnitude and color-color ($U$--$B$
versus $B$--$V$) diagrams for the fields.  We then obtain high
signal-to-noise spectra of these candidates.  With spectroscopy of
cluster WD candidates, it is possible to unambiguously determine if
the objects are WDs.  For the {\it bona fide} WDs, we determine the
effective temperature (\teff) and surface gravity (\logg), and from
those quantities and WD evolutionary models, we derive the cooling age
($\tau_{\mathrm WD}$), mass, and luminosity of each WD.  For those WDs
with cooling ages less than the cluster age and distance moduli
consistent with cluster membership, subtraction of the WD cooling age
from the age of the open cluster results in the lifetime of the
progenitor star, and stellar evolutionary models can then be used to
determine the progenitor star mass.

As part of our ongoing program to identify and spectroscopically
analyze the WD populations of open clusters, the Lick-Arizona White
Dwarf Survey \cite[LAWDS,][]{Williams2004a}, we have obtained imaging
data to identify candidate WDs in NGC 1039 and high signal-to-noise
spectra of WDs in the field of NGC 1039.

In this paper, we assume the solar metallicity value of
\citet[][$Z=0.013$]{2004A&A...417..751A} and the extinction curve of
\citet{1985ApJ...288..618R} with $R_V=3.1$.

\subsection{Previous Studies of NGC~1039}
The first study of WDs in NGC 1039 (M34) was done by
\citet{Anthony-Twarog1982}.  In this study, $\ubv$ photographic plates
were used to obtain photometry for stars in the cluster field and to
select objects with significant blue color excess.  The
color-magnitude sequences for WDs derived by \citet{Sion1977}
and \citet{Koester1979} were compared with these observations.  Four
objects were found to be close to the WD cooling curves used in
color-magnitude and color-color space, and so could be interpreted as
being cluster member WDs.  The number of field WDs of different
apparent $V$ magnitudes expected to appear in the photometry was
calculated using the then-current WDLF \citep{Green1980}.  This number
of expected field WDs ($\sim23$ with $V\leq 21$) was, in fact, larger
than the number of observed blue objects (11 with $\bv\leq 0.4$),
leading to the conclusion that NGC 1039 did not have a significant
number of WDs.

A number of recent studies have been done on NGC 1039.
\citet{Ianna1993} present proper motions to determine which objects
are cluster members.  They determine $(m - M)_0 = 8.28$ and an age of
$\log \tau_{cl} = 8.40$ using $E(B-V)=0.07$.  \citet{Meynet1993}
compare new stellar models to previously obtained data.  They use
$E(B-V)=0.10$ and calculate $(m-M)_0 = 8.34$ and an age of $\log
\tau_{cl} = 8.25$.  \citet{Jones1996}, again using a proper motion
selected sample, use $E(B-V)=0.07$ and obtain $(m-M)_0 = 8.38$ and
$\log \tau_{cl} = 8.30-8.40$.

The most recent study of NGC 1039 is that of \citet{Sarajedini2004}.
These authors combine $U\!BV\!RI$ photometry from the WIYN open
cluster survey with $J\!H\!K$ photometry from 2MASS to obtain accurate
distance estimates.  They assume $E(B-V) = 0.10$ and $[\mbox{Fe/H}] =
+0.07$ and derive $(m-M)_0 = 8.67 \pm 0.07$.  The spectroscopic metal
abundance of NGC 1039 has been determined to be slightly super-solar.  
\citet{Schuler2003} find $[\mbox{Fe/H}] = +0.07
\pm 0.04$, and new work from the WIYN Open Cluster Study finds
$[\mbox{Fe/H}]\approx +0.08$ (A.\ Steinhauer, private communication, 2008).  
In this paper, we assume solar metallicity for the cluster.

\section{Photometric Observations \& Analysis}

\subsection{Photometric Measurements and Calibration}

$\ubv$ imaging of a field centered on NGC 1039, as well as of two
fields $25'$ to the north and south of the cluster center, was
obtained in 2004 September and 2005 December with the KPNO 4m MOSAIC
camera.  Exposures of 5, 30, and $3\times 120$ s were obtained at
the northern field in both $B$ and $V$ and at the central field in
$V$; exposures of 5, 30, and $3 \times 150$ s were taken at the
central field in $B$.  We obtained exposures of 5 and 120 s in $B$
and $V$ at the southern field.  Exposures of 5, 30, 120, and
$3\times900$ s were obtained in $U$ at the central field; all of
these but the 30 s exposure were obtained in the northern field.
The southern field was observed with exposures of 15, 120, and $3
\times 600$ s in $U$.  The northern and southern fields are
positioned at very nearly the same right ascension (R.A.) as the central field.  Each
flanking field overlaps the central field by $\sim 11\arcmin$ on
either its northern or southern end.  Seeing ranged between $0\farcs
8$ and $1\farcs 0$ in the $V$-band in 2004 September and was $\sim
1\farcs 7$ in the $B$-band in 2005 December.

We reduced the data using the IRAF $mscred$ mosaic reduction
package.  The data were bias subtracted, trimmed, flat-fielded and
projected onto the tangent plane.  Object detection and aperture
photometry were performed using the DAOPHOT II program
\citep{Stetson1987}; aperture corrections were determined and applied
using the program DAOGROW \citep{Stetson1990}.

Star-galaxy separation was performed by comparing the $V$-band flux in
two successively larger apertures for each object: an inner aperture
with a radius equal to the full-width half-maximum (FWHM) of the
$V$-band point spread function ($\approx 0\farcs\, 8$) and an outer
aperture with a radius of twice the FWHM.  Stars were observed to form a
tight sequence with a constant difference in aperture magnitudes of
$\approx 0.3$ mag for $V\lesssim 23$; we therefore identified objects
with aperture magnitude differences $\leq 0.5$ and $V\lesssim 23$ as
point sources.  We consider only these point sources in the subsequent
analysis.

As the MOSAIC images of NGC 1039 were taken under non-photometric
conditions, we calibrated these images using previously obtained
$\ubv$ imaging of the central portion of NGC 1039 taken with the
Nickel 1-m telescope at Lick Observatory on 2002 July 6. These images
were reduced and aperture photometry was determined for each star.
The Nickel imaging was calibrated using standard stars from
\citet{Landolt1992}. The stars with calibrated Nickel photometry were
then used as secondary standard stars for calibrating the MOSAIC
imaging.

Calibrations were obtained by solving the following transformation
equations for both the MOSAIC and Nickel imaging:
\begin{eqnarray}
        \scriptstyle u & \scriptstyle =& \scriptstyle U + 2.5 \log t_{exp} + A_0 + A_1 (X-1.25) + A_2 (U - B)
        + A_3 (U - B)^2 \\
        \scriptstyle b & \scriptstyle =& \scriptstyle B + 2.5 \log t_{exp} + B_0 + B_1 (X-1.25) + B_2 (B - V) \\
        \scriptstyle v & \scriptstyle =& \scriptstyle V + 2.5 \log t_{exp} + C_0 + C_1 (X-1.25) + C_2 (B - V),
\end{eqnarray}
where $u$, $b$, $v$ are total instrumental magnitudes; $U$, $B$, $V$
are standard magnitudes; $t_{exp}$ is the exposure time; and $X$ is the
airmass.

Zero points and color terms were calculated for the MOSAIC data; as
local secondary standards were used, the airmass terms were not
necessary.  Values for the transformation coefficients are given in
Table \ref{tab.coef}.  The $B$ and $V$ color terms are consistent with
average values determined by \citet{Massey2006} for the MOSAIC camera;
the $U$-band color terms are not directly comparable because we have
chosen a quadratic formula.

We compared our calibrated MOSAIC photometry of 46 objects with that
of \citet{Jones1996}.  There is no significant trend in the magnitude
of $\bv$ color offset with magnitude or color between the two systems.
The mean offset in $V$ magnitude is 0.021, with a dispersion of 0.048.
The mean offset in $\bv$ color is 0.008, with a dispersion of 0.050.

The left most panels of Figures~\ref{fig.cmd} and \ref{fig.uvcmd} show
our final calibrated photometry for the cluster in $B-V$, $V$ and
$U-V$, $V$ color-magnitude diagrams.  Note that sources with $V
\lesssim 13$ are saturated in our photometry.  \citet{Jones1996}
photometry is overplotted in the second left panel in
Figure~\ref{fig.cmd} for stars with a high probability of cluster
membership ($> 67\%$, where probability is determined from their
proper motion study).

\subsection{WD Candidate Selection}

We select WD candidates using photometric criteria based on synthetic
evolutionary sequences of WDs kindly provided by P.\
Bergeron\footnote{http://www.astro.umontreal.ca/\~{}bergeron/CoolingModels/};
these sequences represent an extension of those published in
\citet{Bergeron1995}, including improved color calculations
\citep{Holberg2006} and the more recent WD evolutionary calculations
of \citet{Fontaine2001}.

\begin{figure*}
\includegraphics[scale=0.85]{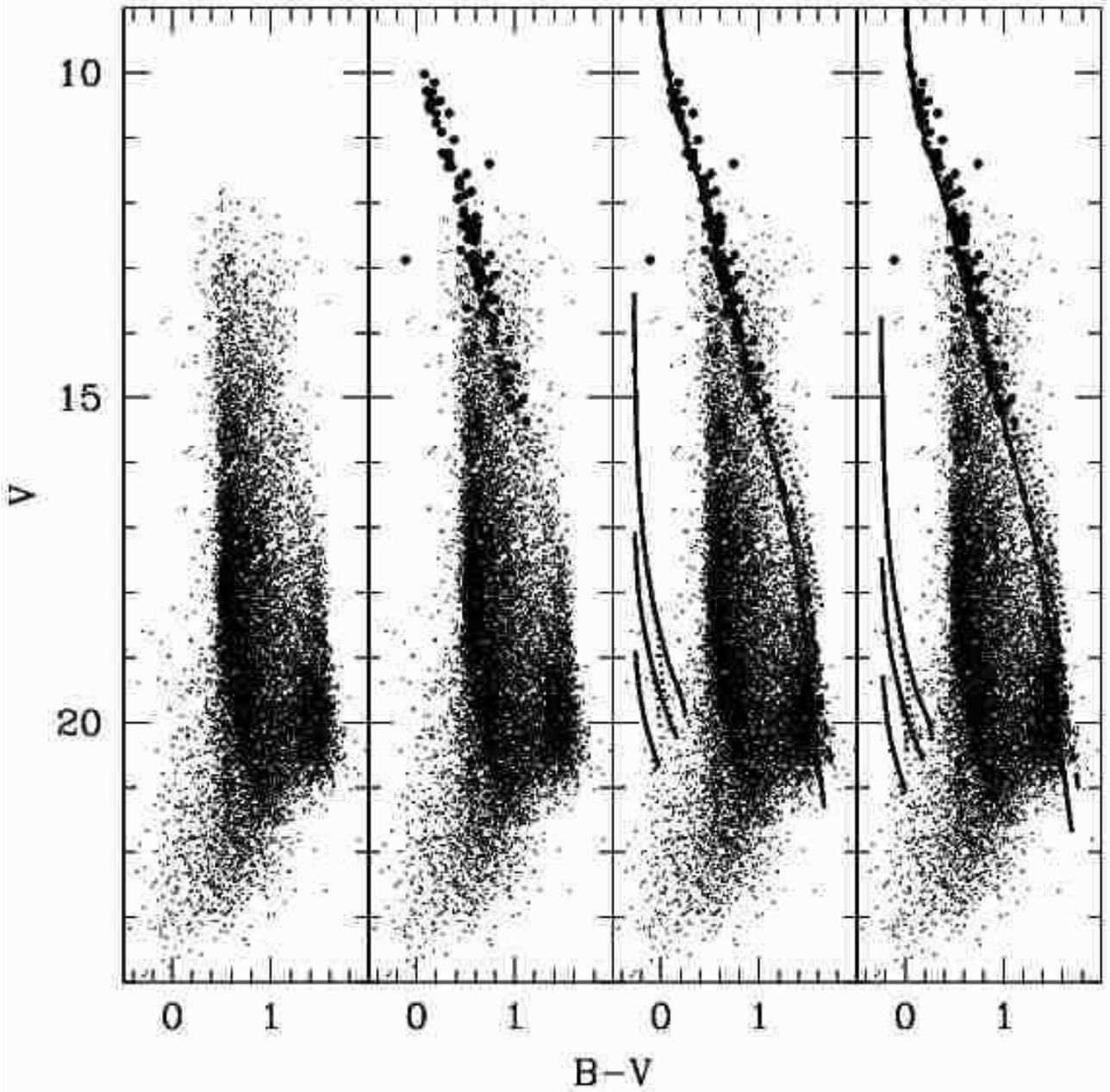}
\caption{$\bv,\ V$ color-magnitude diagram for NGC 1039.
  Left: our photometry alone; note that sources with $V
  \lesssim 13$ are saturated. Second left: our photometry
  plus that of cluster members from \citet{Jones1996}. Second right: combined
  photometry with isochrones and cooling curves overplotted using the
  \citet{Jones1996} determination of the cluster $(m-M)_V$.
  Right: combined photometry with isochrones and cooling
  curves adjusted to the cluster $(m-M)_V$ derived by
  \citet{Sarajedini2004}.  In all panels, small dots indicate our
  photometry and larger dots show high-probability proper motion cluster
  members from \citet{Jones1996}.  The main-sequence isochrones are
  a $Z=0.013$, $\log \tau_{cl} = 8.35$ Padova isochrone interpolated
  from \citet[solid]{Girardi2002}, a $Z=0.013$, $\log \tau_{cl} = 8.35$
  Yonsei-Yale isochrone \citep[dashed]{Demarque2004}, and the fiducial
  main sequence of NGC 2168 from \citet[dotted]{vonHippel2002}
  adjusted to the metallicity of NGC 1039.  Cooling curves for cluster
  WDs with cooling times less than 300 Myr and masses of $0.4
  M_{\odot}$ (top), $0.8 M_{\odot}$ (middle) and $1.2 M_{\odot}$
  (bottom) are shown for DAs (solid) and DBs (dotted).
  \label{fig.cmd}}

\end{figure*}

\begin{figure*}
\plotone{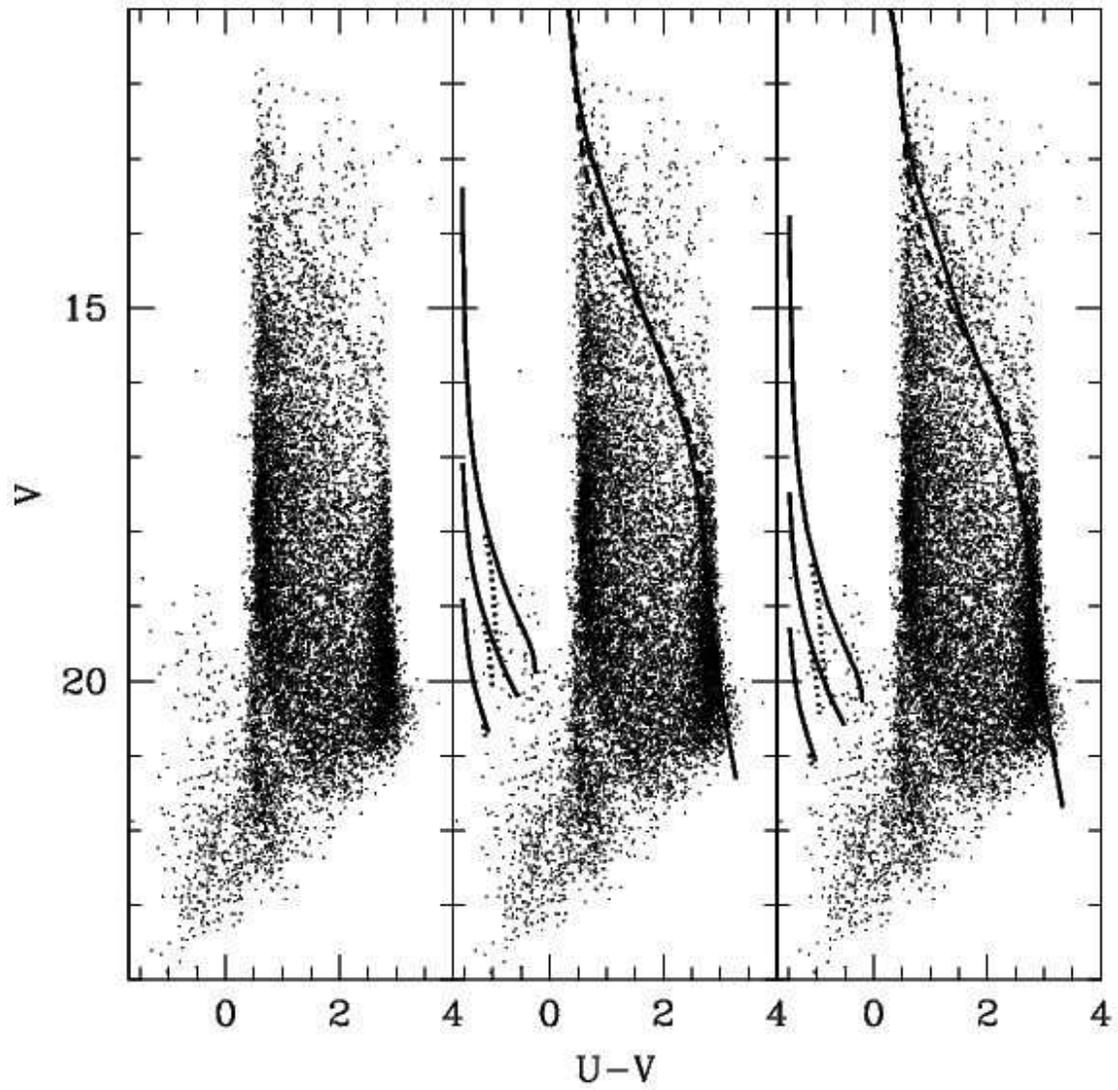}
\caption{$U$--$V$, $V$ color-magnitude diagram for NGC 1039.  Symbols
  are as in Figure \ref{fig.cmd}. 
Left: our photometry alone.  Middle: our photometry
  with isochrones and cooling curves overplotted using the
  \citet{Jones1996} determination of the cluster
  $(m-M)_V$.  Right: our photometry with isochrones and
  cooling curves adjusted to the cluster $(m-M)_V$ derived by
  \citet{Sarajedini2004}. \label{fig.uvcmd}}
\end{figure*}

The second right panel of Figure \ref{fig.cmd} and the middle panel
of Figure \ref{fig.uvcmd} show isochrones and WD cooling curves adjusted to
the cluster's $(m-M)_{V}$ and \ebv using cluster parameters from
\citet{Jones1996}.  The right-hand panels of both figures show the
same isochrones and WD cooling curves, adjusted using cluster
parameters from \citet{Sarajedini2004}.  The solid isochrones are
Padova models for $\log \tau_{cl} = 8.35$ \citep{Girardi2002},
interpolated for $Z=0.013$.  The dashed isochrones are Yonsei-Yale
$\log \tau_{cl} = 8.35$, $Z=0.013$ models interpolated from
solar-scaled abundance models using the code provided by the Yonsei-Yale
Group \citep{Demarque2004}.  The dotted isochrone in
Figure \ref{fig.cmd} is the fiducial main sequence of the sub-solar
metallicity cluster NGC 2168 given in \citet{vonHippel2002}, corrected
to the metallicity of NGC 1039 using the method described in
\citet{Pinsonneault1998}.  The cooling curves for DA and DB WDs of
different masses are also shown.
 
It is evident from Figure \ref{fig.cmd} that our photometry matches
the photometry in the literature acceptably well, and that modeled
isochrones follow the $\bv$ versus $V$ main sequence in our data. The
best fit of the observed main-sequence to model isochrones is obtained
when we adopt the $E(B-V)$ and $(m-M)_{V}$ values from
\citet{Sarajedini2004}; model isochrones shifted to the distance
modulus and reddening from \citet{Jones1996} fail to match the
observed main sequences. We therefore simply adopt the values of
$E(B-V)$ and $(m-M)_{V}$ given in \citet{Sarajedini2004} for the
remainder of our analysis.

\begin{figure*}
\plotone{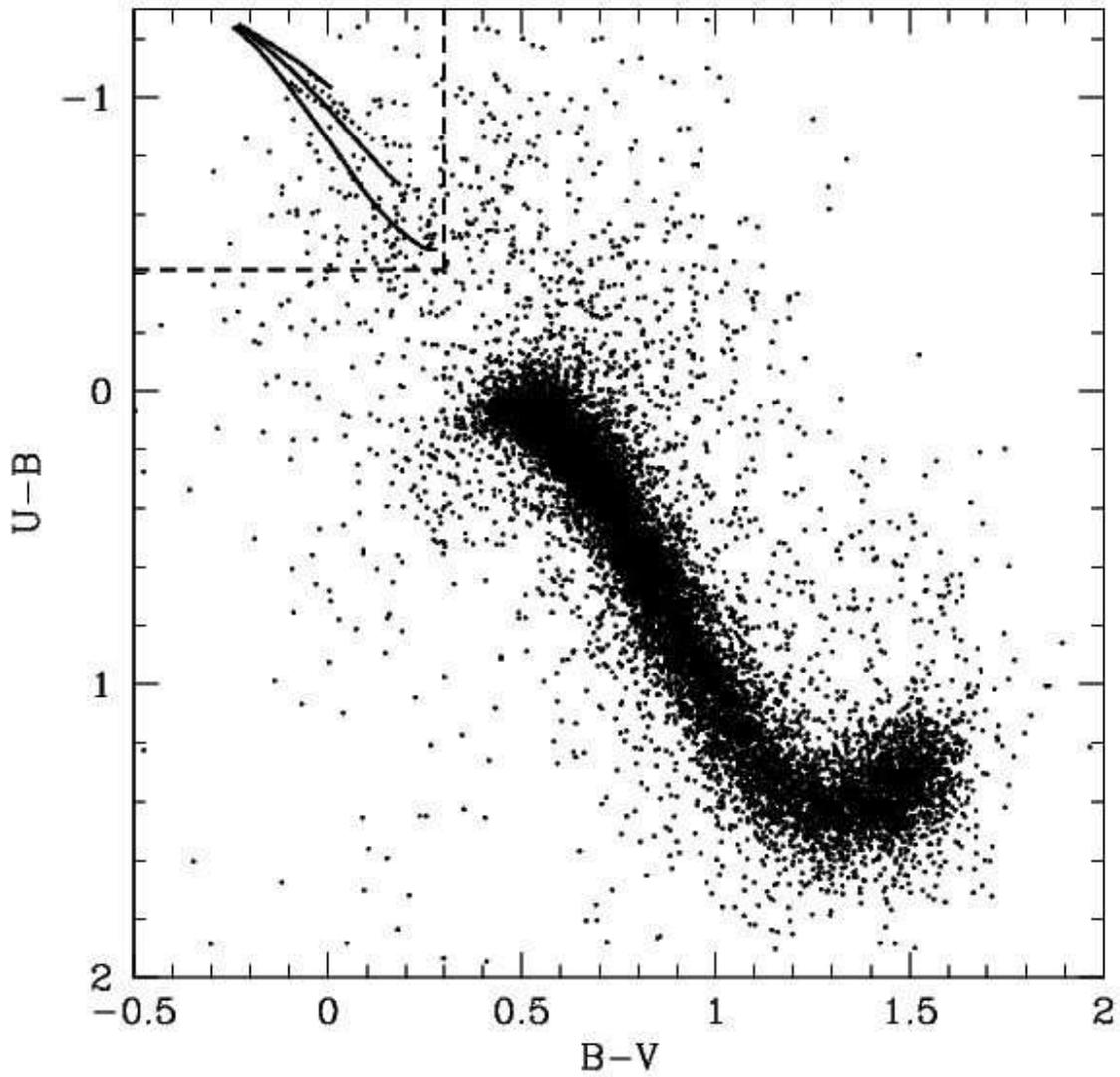}
\caption{$\bv, U$--$B$ color-color diagram for NGC 1039.  Cooling
curves for cluster WDs with cooling times less than 300 Myr and masses
of $1.2 M_{\odot}$ (top), $0.8 M_{\odot}$ (middle), and $0.4 M_{\odot}$
(bottom) are shown for DAs (solid) and DBs (dotted).  Vertical and
horizontal dashed lines mark the boundaries of the WD candidate
selection region.  \label{fig.colorcolor}}
\end{figure*}

As seen in Figure \ref{fig.uvcmd}, not one of the plotted isochrones
appears to match our main sequence photometry in the fainter parts of
the $U-V$ versus $V$ color-magnitude diagram well; the observed main
sequence is significantly redder than any theoretical isochrone for
$V\gtrsim 14$ ($M_V\gtrsim 5$).  The cause of this discrepancy is
unclear. Given the agreement between models and observations in the
$B-V,\, V$ plane, the problem is likely either shortcomings in $U$-band
model spectra for cool stars or calibration issues in our $U$-band
photometry.  Given the agreement between our WD photometry
and photometric models (see \S\ref{sec.initmass}), the latter seems
unlikely.  Figure \ref{fig.colorcolor} shows a color-color diagram
with our photometry in the field of the cluster.

\begin{figure*}
\plotone{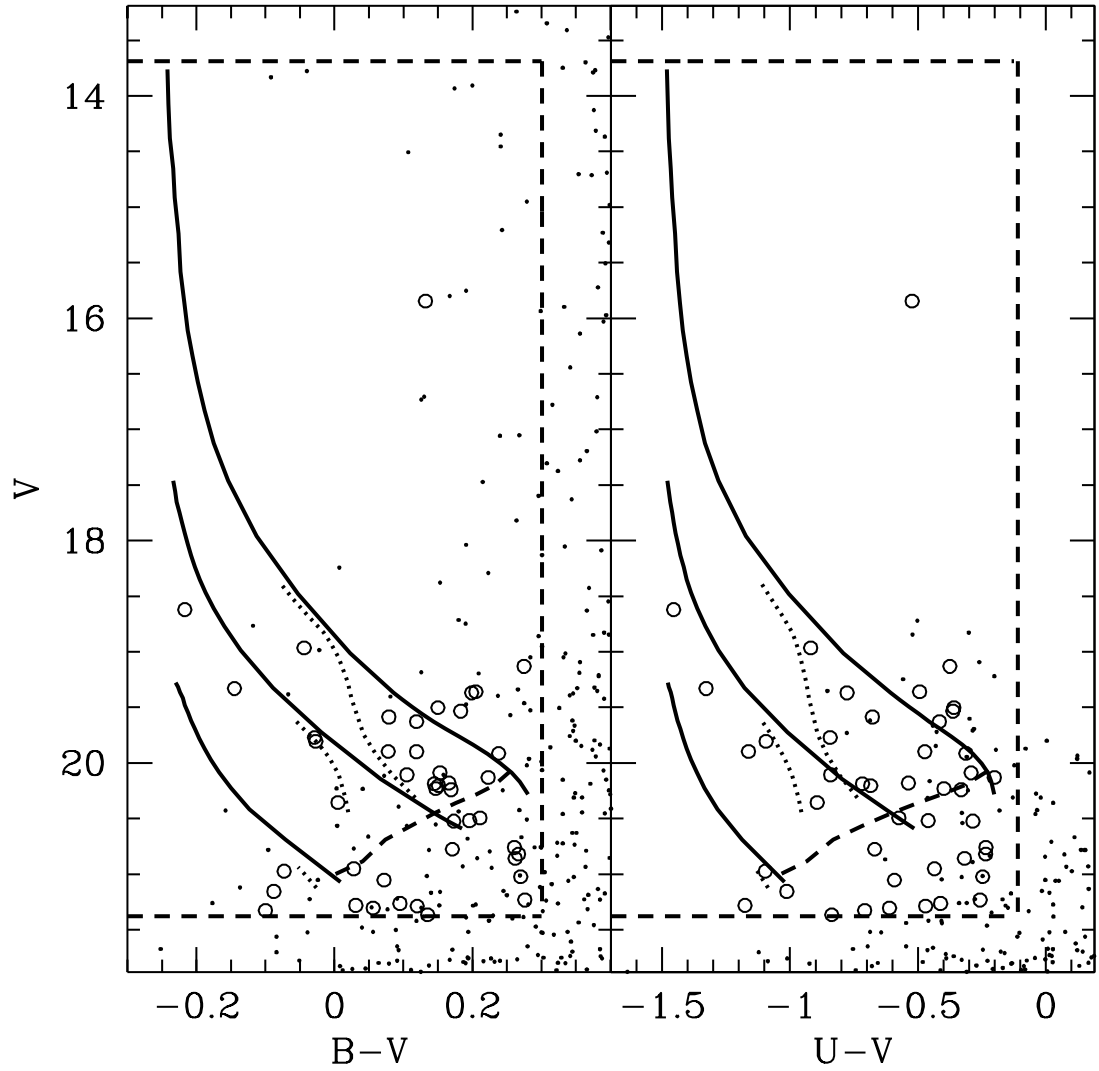}
\caption{Lower left-hand regions of the color-magnitude diagrams for
NGC 1039.  Cooling curves are described in Figures~\ref{fig.cmd} and
\ref{fig.uvcmd}.  Dashed lines at the faint ends of the cooling curves
mark the colors and $V$ magnitudes of WDs which have been cooling for
$\log \tau = 8.35$ ($\sim 225$ Myr).  Vertical and horizontal dashed lines mark
boundaries of the WD candidate selection region.  Open circles
indicate objects that meet all of our WD selection criteria.
\label{fig.selcmd}}
\end{figure*}

\begin{figure*}
\plotone{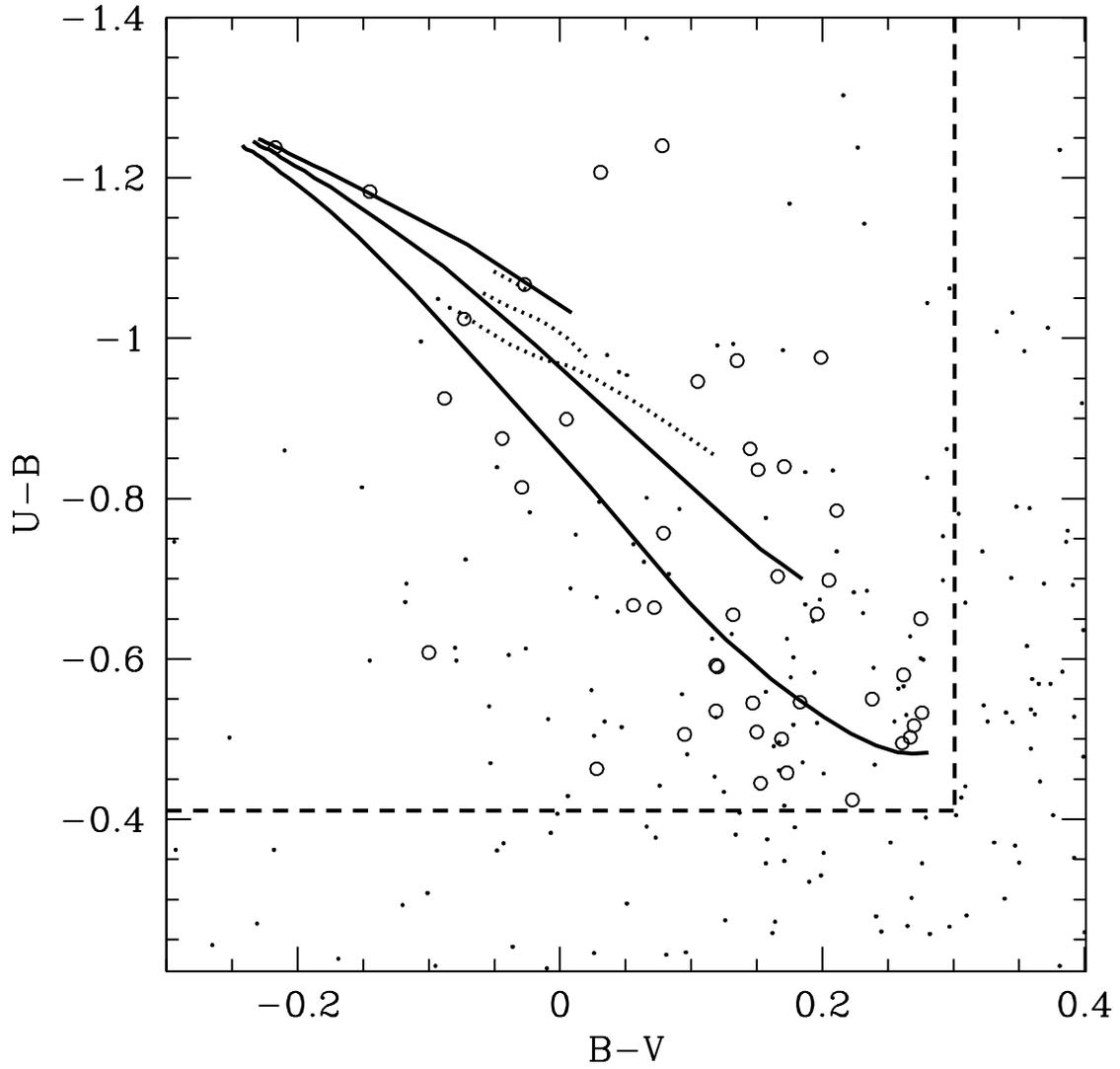}
\caption{Upper left-hand region of the color-color diagram for NGC
1039.  Cooling curves are described in Figure~\ref{fig.colorcolor}.
Vertical and horizontal dashed lines mark boundaries of the WD
candidate selection region.  Open circles indicate objects that meet
all of our WD selection criteria. \label{fig.selcolor}}
\end{figure*}

Figures~\ref{fig.selcmd} and \ref{fig.selcolor} show the sections of
the color-magnitude diagrams and color-color plot surrounding the WD
cooling sequences with our photometry of the field of the cluster.
Cooling curves are shown as in Figures~\ref{fig.cmd}, 
\ref{fig.uvcmd}, and \ref{fig.colorcolor}.  The boundaries of our selection regions for cluster
WD candidates are shown with dashed lines.  The $V$-magnitude limits
for WD candidate selection are $13.69 \leq V \leq 21.38$.  We also
required that all WD candidates have $\bv \leq 0.30$ and $U-B \leq
-0.41$.  These boundaries include all photometric models for DA WDs in
the mass range $0.4 M_{\odot} < M < 1.2 M_{\odot}$ with $\tau_{\mathrm WD}
\lesssim 8.48$, adjusted for the adopted distance and reddening of the
cluster.  Objects which meet all of these criteria were selected as
possible WD candidates for spectroscopic followup.  They are marked in 
Figures~\ref{fig.selcmd} and \ref{fig.selcolor} with open
circles.  Table \ref{tab2.phot} lists $\ubv$ photometry of all 44
selected WD candidates.  Seven objects from \citet{Anthony-Twarog1982}
are recovered.  See our \S\ref{discussion.AT} for a more detailed
description of these objects.

\section{Spectroscopic Observations \& Analysis}

\subsection{Spectroscopy of Candidate WDs}

Spectroscopic observations of selected WD candidates were obtained
over several observing runs between UT 2001 August 22 and 2007 January 20
with the blue camera of the LRIS spectrograph \citep{Cohen1994} on the
Keck 1 10-m telescope \citep{Oke1995}. The 2001 August observations
used the initial LRIS-B engineering-grade $2048 \times 2048$ SITe CCD;
all other observations used the New Blue Camera, consisting of two 2k
$\times$ 4k Marconi CCDs.  The 2001 observations therefore have lower
sensitivity in the blue.

A 1\arcsec-wide longslit at parallactic angle was used with the 400 l
mm$^{-1}$, 3400\AA~blaze grism for a resulting spectroscopic
resolution of $\sim 6$ \AA.  The spectra were reduced using the
$onedspec$ package in IRAF.  Overscan regions were used to
subtract the amplifier bias, and cosmic rays were removed from the
two-dimensional spectra using the L.A. Cosmic Laplacian cosmic ray
rejection routine \citep{vandokkum2001}.  We then co-added
any multiple exposures of an individual object and extracted
the one-dimensional spectrum.  We applied a wavelength solution
derived from HgCdZn arc lamp spectra.  We applied a relative flux
calibration determined from longslit spectra of spectrophotometric
standard stars.

\begin{figure*}
\plotone{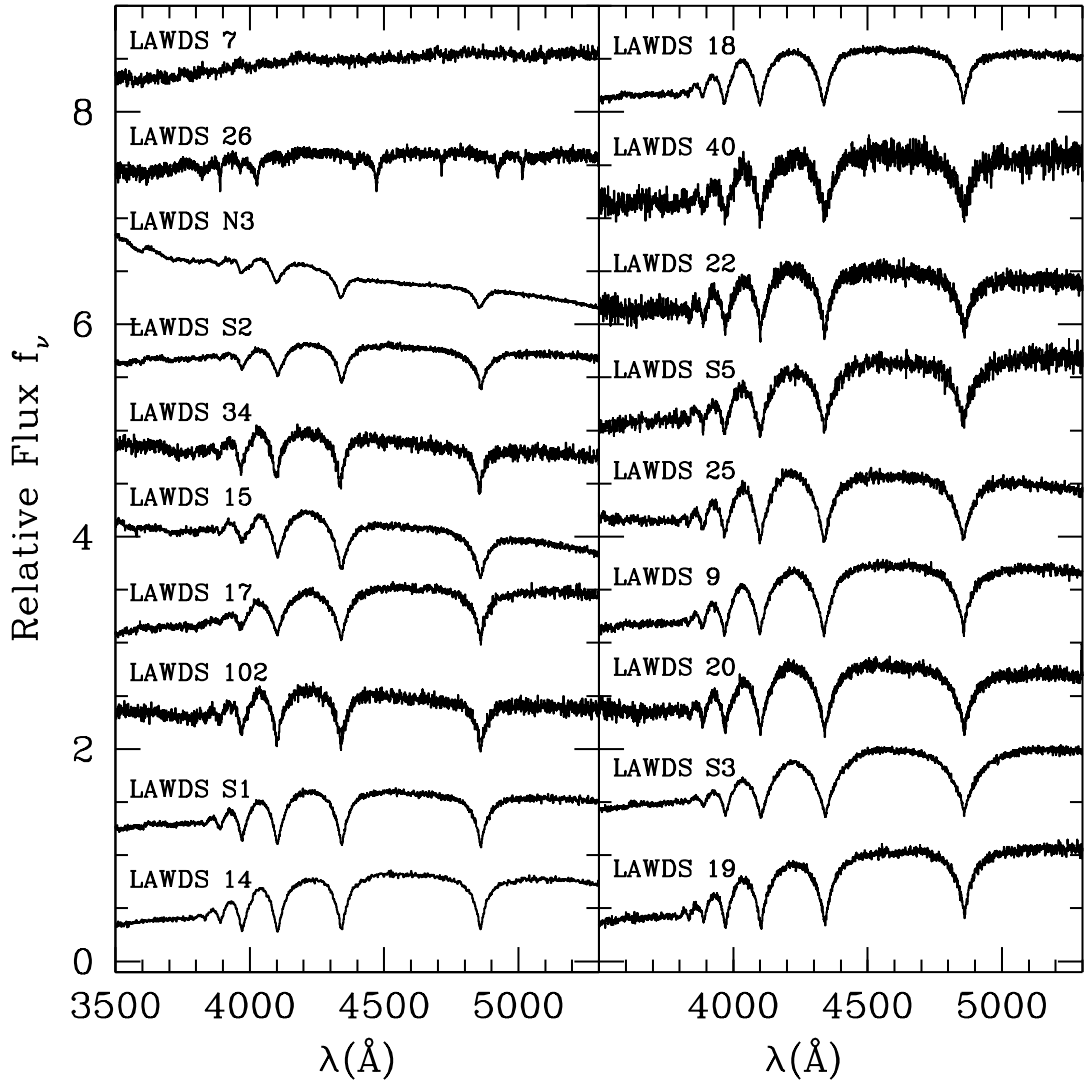}
\caption{Spectra of WDs in the field of NGC 1039.  The spectra of
the DB (LAWDS 26) and DC (LAWDS 7) WDs are shown at the top of the
left-hand panel, followed by the spectra of the DA WDs in order of
decreasing WD \teff, from top to bottom and continuing through the
right-hand panel.  \label{fig.allspecs}}
\end{figure*}

Spectra of WDs are shown in Figure~\ref{fig.allspecs}.  Spectroscopic
identification of each WD candidate is given in Table~\ref{tab2.phot}.
The major non-WD contaminants in the sample are active galactic
nuclei (AGNs).  To determine redshifts for the latter objects, we
cross-correlated the spectra with the Sloan Digital Sky Survey (SDSS)
composite quasi-stellar object (QSO) spectrum of \citet{VandenBerk2001} as described in
\citet{Williams2004b}.

\subsection{WD Parameter Determination}

 \teff and \logg were determined for each WD using simultaneous
Balmer-line fitting \citep{Bergeron1992}, applied as described in
\citet{Williams2007}.  The model spectra used for the fits are from
Koester's model grids; details of the input physics and methods can be
found in \citet{Koester2001} and references therein.  DA evolutionary
models were provided freely by P.\ Bergeron, and consist of models of
\citet{Wood1995} (for $\teff > 30,000$K) and \citet{Fontaine2001} (for
$\teff\leq 30,000$ K), with ``thick'' hydrogen layers ($q_H=10^{-4}$)
and mixed C/O composition.  These models were used to calculate the
mass ($M_{\mathrm WD}$) and cooling age ($\tau_{\mathrm WD}$) of each
WD.

We consider three primary sources of error in these parameters:
internal fitting errors (what is the expected distribution in the fit values
if we fit multiple observations of the same star with the same
signal to noise?), external fitting errors (what is the expected
distribution in the fit values if others were to observe the same star
and derive independent fits?), and systematic error due to the cluster
age, which affects all WDs in the same sense, and so should not be
considered as a random error to be added in quadrature. We discuss the
two fitting errors here, and the systematic errors in
\S\ref{sec.initmass} and \S\ref{discussion.ifmr}.

Internal fitting errors were determined empirically.  The
noise measured for each spectrum was added to the best-fitting model
spectrum convolved with the instrumental response.  These simulated
spectra were then fitted by the same method; nine iterations were used to
calculate the scatter in \teff, \logg, $M_{\mathrm WD}$, and
$\tau_{\mathrm WD}$.  The errors thus determined ($\sigma_{\teff, {\rm
  int}}$ and $\sigma_{\logg, {\rm int}}$) represent \emph{only}
random errors due to observations and spectral fitting.

External fitting errors can be estimated for our data from the
comparison of field WD parameters performed in \citet{Williams2007}.
This analysis found that the scatter in WD parameters due to the use
of different instrumentation, spectral fitting routines, and model
atmospheres is $\sigma(\teff)\approx 1100$K and $\sigma(\logg)\approx
0.12$.  These are added in quadrature with the internal fitting errors
to produce the adopted values for $\sigma_{\teff, {\rm tot}}$ and
$\sigma_{\logg, {\rm tot}}$ for each WD.

\begin{figure*}
\includegraphics[scale=0.85  ]{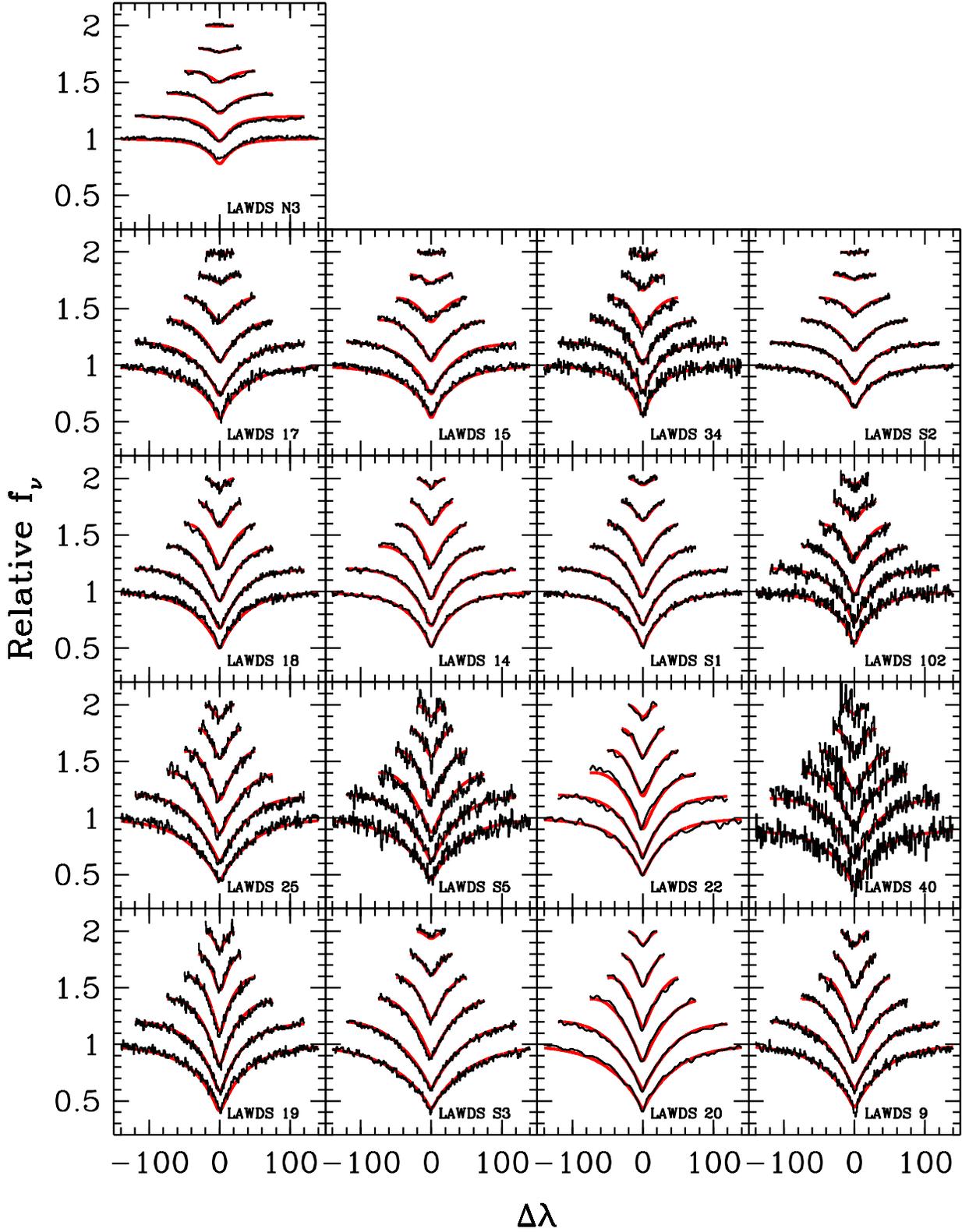}      
\caption{Balmer-line profiles (light histograms) and best-fit models
  (heavy curves) for NGC 1039 DA WD spectra.  The abscissa shows the
  offset in Angstroms from the center of each Balmer line.  From
  bottom to top in each panel, the H$\beta$ through H9 line regions
  are plotted.  Each section of spectrum has been normalized to the
  pseudocontinuum just outside the plotted region and offset
  arbitrarily in the vertical direction.  (A color version of this figure is
  available in the online journal.)
\label{fig.allfits}}
\end{figure*}

The atmospheric fits and derived WD masses and ages are given in
Table~\ref{tab.spec}.  The Balmer-line fits are shown in
Figure~\ref{fig.allfits}.  A distance modulus for each WD was derived
by comparing the observed $V$ magnitude to the absolute magnitude
$M_V$ calculated from the best-fitting model atmosphere and the
appropriate WD cooling model.

\section{WD Cluster Member Identification}

\subsection{WD Cluster Membership Determination}
\begin{figure*}
\plotone{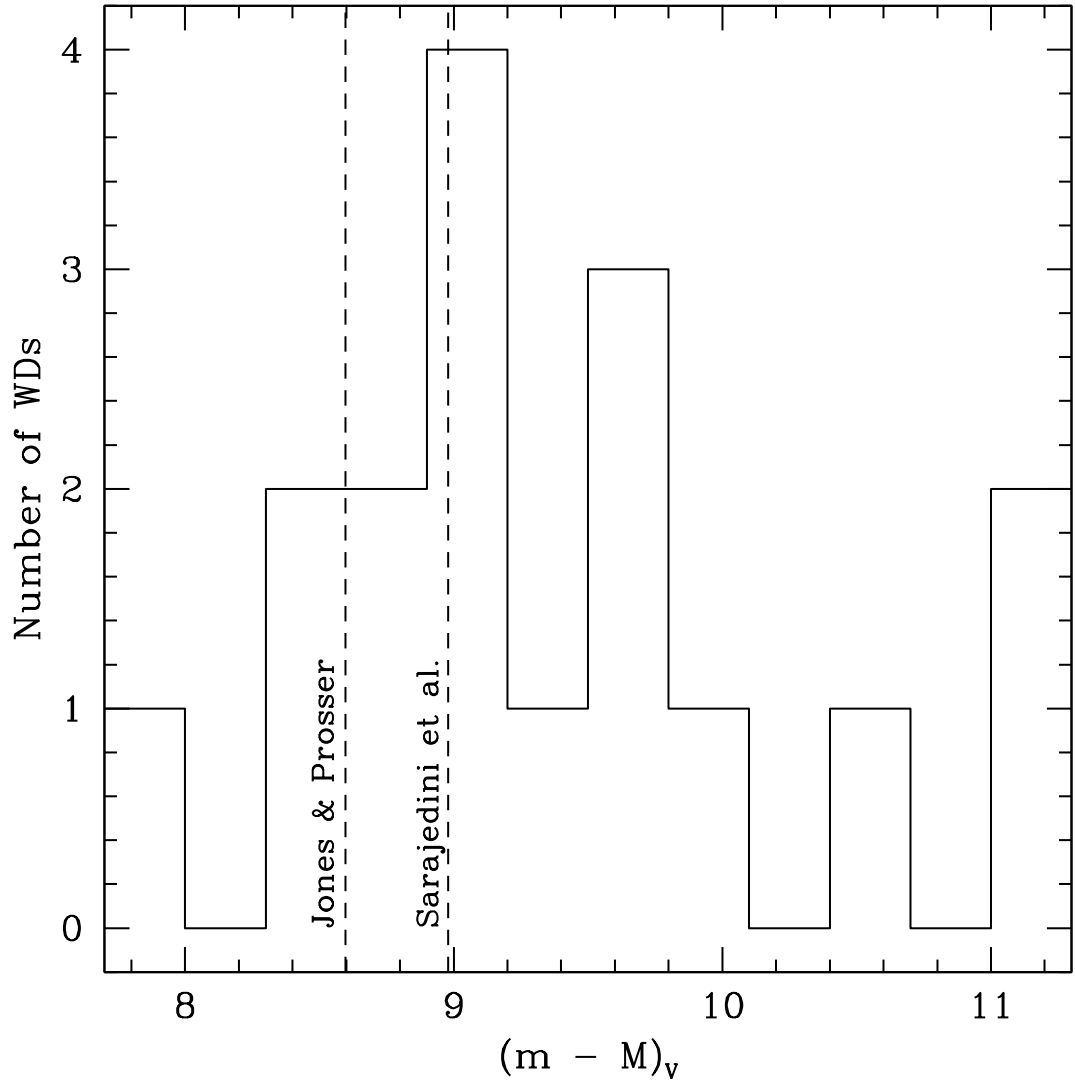}
\caption{Histogram for all spectroscopically confirmed DA WDs in
$(m-M)_{V}$ bins of width 0.3 mag.  The \citet{Jones1996} and
\citet{Sarajedini2004} determinations of $(m-M)_{V}$ for NGC 1039 are
marked with dashed lines.  \label{fig.dmhist}}
\end{figure*}

We consider two criteria when determining whether a WD is a potential
member of NGC 1039. (1) Its $(m-M)_{V}$ must be between the two
determinations of the cluster $(m-M)_{V}$ made by \citet{Jones1996}
and \citet{Sarajedini2004} and (2) its cooling age must be less than
the cluster age.  Figure~\ref{fig.dmhist} shows $(m-M)_{V}$ for each
observed DA WD in a histogram.  The \citet{Jones1996} and
\citet{Sarajedini2004} distance moduli determinations are marked with
dashed lines.  We select five DA WDs to be potential members of NGC
1039 based on this distance modulus and age selection. Those WDs with
distance moduli up to 0.75 $V$ magnitudes brighter than the
\citet{Jones1996} cluster distance modulus are in the appropriate
brightness range to be in double-degenerate binary systems in the
cluster.  There is one such candidate among our DA WDs which is
younger than the cluster.  Table~\ref{tab.masses} lists all selected
cluster member candidates.
 
 \subsection{Progenitor Mass Calculation\label{sec.initmass}}
 The progenitor mass (initial mass) for each candidate cluster member
 WD was calculated by first subtracting the WD cooling age
 ($\tau_{\mathrm WD}$) from the cluster age.  The resulting age
 difference is the total lifetime of the progenitor star from the
 zero-age main sequence (ZAMS) to the thermally pulsing asymptotic
 giant branch (AGB) phase; the amount of time between the AGB phase and
 the entrance of the star into the WD cooling track is assumed to be
 negligible.  We then calculated the ZAMS mass of a star with that
 total lifetime from the solar-metallicity stellar evolutionary models
 of \citet{Girardi2000} and \citet{Bertelli1994}.  The progenitor mass
 for each candidate cluster member DA WD is given in
 Table~\ref{tab.masses} for three ages of NGC 1039 spanning the range
 of recently published ages: $\log\tau_{cl}=8.30\,(\approx$ 200 Myr),
 $\log\tau_{cl}=8.35\,(\approx$ 225 Myr), and
 $\log\tau_{cl}=8.40$ \citep[$\approx$ 250 Myr;][]{Jones1996}.  Errors in
 the initial masses are $1\sigma$ variations resulting from the
 combined internal and external fitting errors in \logg ~and \teff.

\begin{figure*}
\plotone{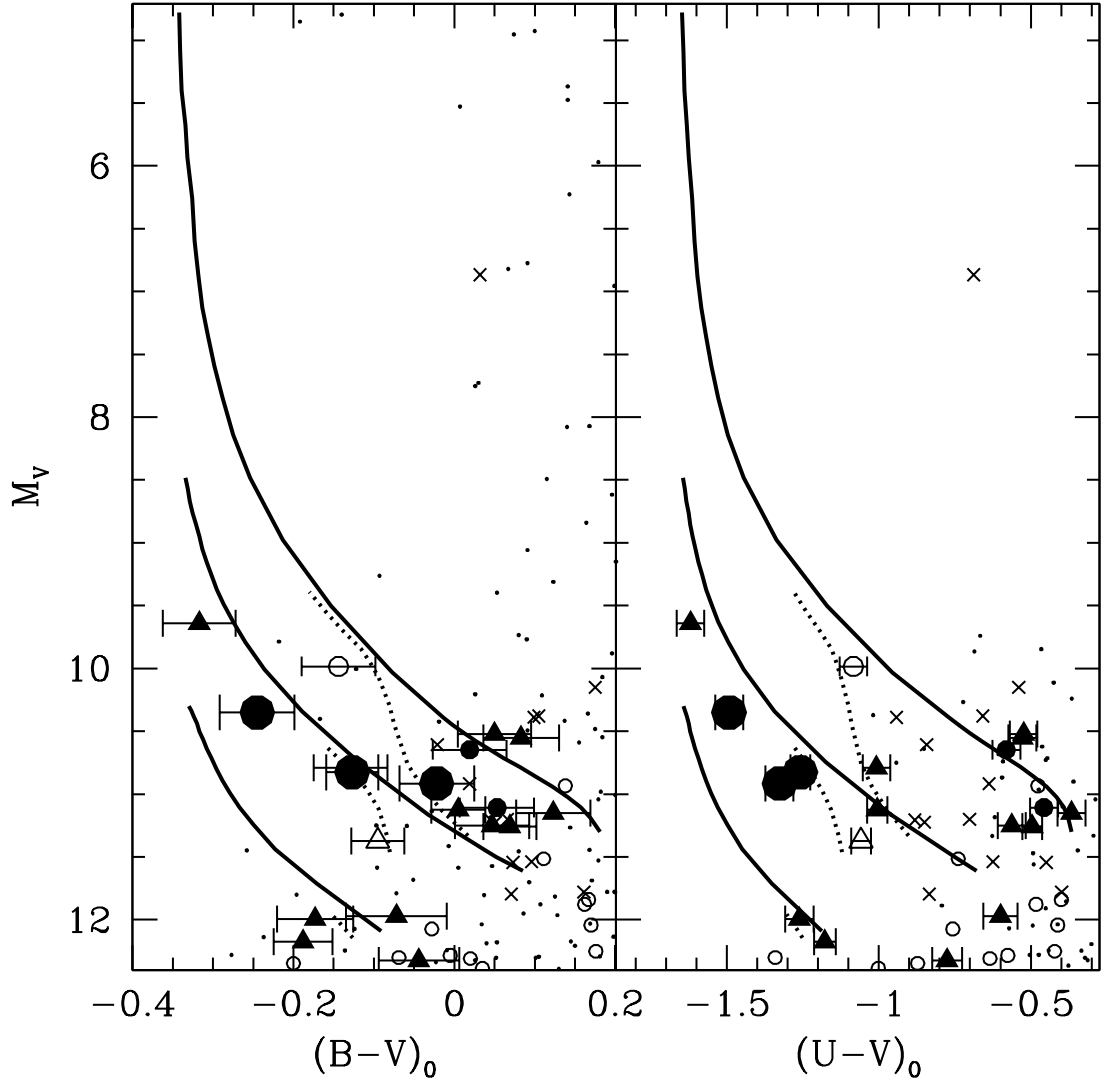}
\caption{WD candidate selection region of the color-magnitude diagrams
for NGC 1039.  Cooling curves are described in Figures~\ref{fig.cmd}
and \ref{fig.uvcmd}.  The symbols indicate high mass ($> 0.8
M_{\odot}$) cluster member WDs (large filled circles), lower mass
cluster member WDs (intermediate filled circles), the candidate
double-degenerate WD (intermediate open circle), the DB (membership 
undetermined; open triangle), non-cluster member
WDs (small filled triangles), non-WDs (crosses), unobserved WD
candidates (small open circles), and stars not meeting our selection
criteria (small points).   Horizontal error
bars show $1\sigma$ uncertainties in de-reddened color.  $1\sigma$
uncertainties in $M_{V}$ for WDs are smaller than the plotted point
sizes.  \label{fig.resultscmd}}
\end{figure*}

\begin{figure*}
\plotone{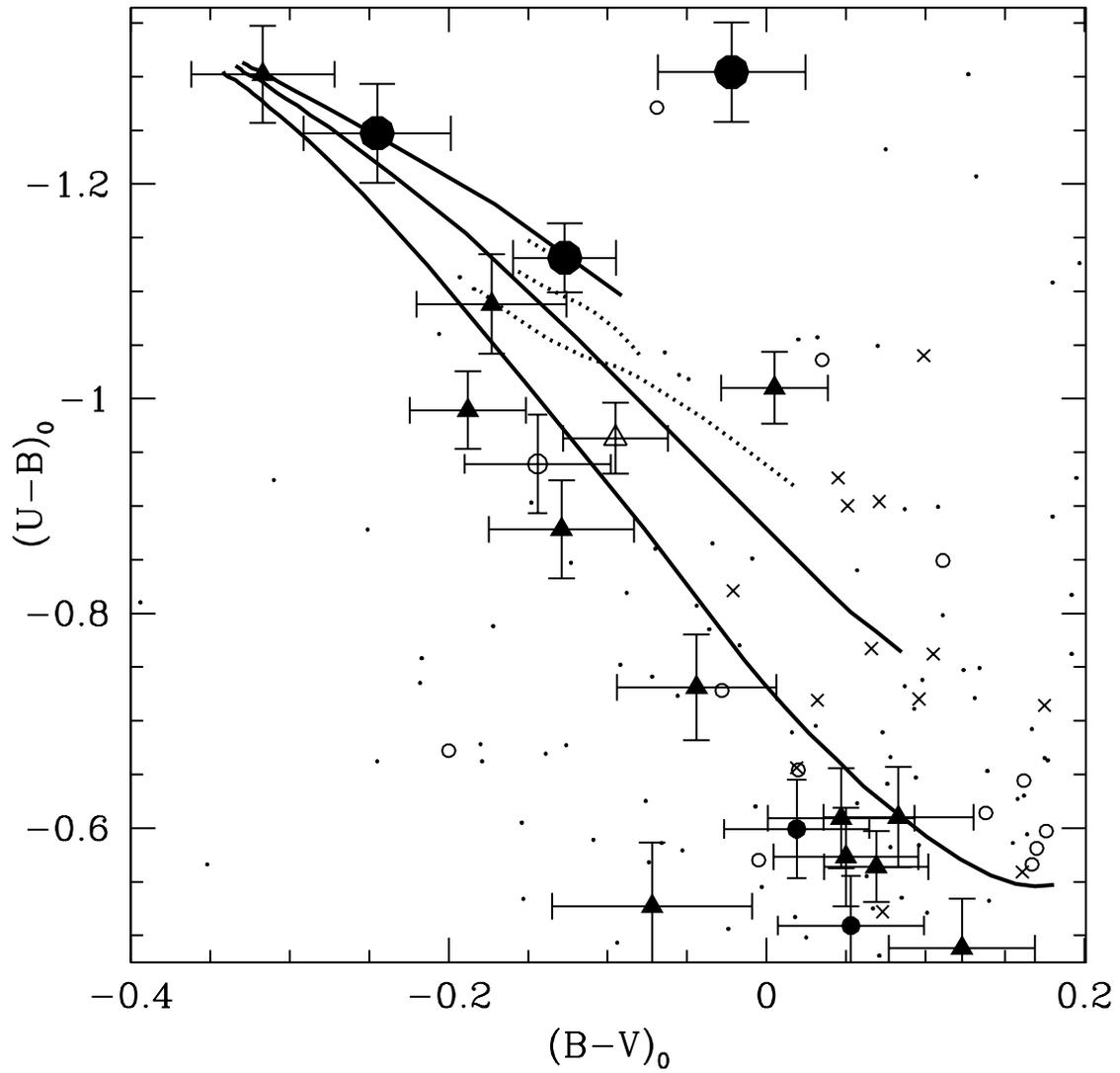}
\caption{WD candidate selection region of the color-color diagram for
NGC 1039.  Cooling curves are described in
Figure~\ref{fig.colorcolor}; symbols are as in Figure
\ref{fig.resultscmd}.
Error bars show $1\sigma$ uncertainties in de-reddened color.
\label{fig.resultscolorcolor}}
\end{figure*}

Figures \ref{fig.resultscmd} and \ref{fig.resultscolorcolor} show our
WD candidate selection region of color-magnitude and color-color space
with objects spectroscopically identified.  Of particular note are the
filled and open circles.  Large filled circles represent high mass ($
> 0.8 M_{\odot}$) DA WD cluster members (LAWDS 15, 17, and S2; see the next
section).  These lie quite close to the plotted $0.8 M_{\odot}$ DA WD
cooling curves in both color-magnitude diagrams in Figure \ref{fig.resultscmd}.  
Two of these lie close to the plotted cooling curves in the color-color diagram 
(Figure \ref{fig.resultscolorcolor}); one of them (LAWDS 17) lies to the right of the 
cooling curves.  The
intermediate-sized open circle represents the object selected as a
candidate double-degenerate cluster member (LAWDS S1).
Intermediate-sized filled circles represent lower mass DA WDs selected
to be cluster members (LAWDS 9 and 20).

\subsection{Field WD Contamination\label{discussion.contamination}}
We select WDs for cluster membership based only on their distance and
cooling time, yet it is quite possible for field WDs to satisfy these
criteria as well.  We estimate the probability of finding field WDs
meeting our selection criteria in the WD selection volume using the
\citet{Harris2006} luminosity function.  In the $36\arcmin \times
86\arcmin$ MOSAIC field, $\sim 2.4$ field WDs with $M_{V} \leq 11.8$
(the faintest $M_V$ of the confirmed WDs) are expected.

Three of our WDs selected as cluster members are $\sim 0.8 M_{\odot}$,
roughly the value that would be expected for NGC 1039 WDs based on the
cluster age and previously published empirical IFMRs.  (Another WD of
similar mass, LAWDS S3, is too close [$(m-M)_{V} = 7.72$] to be
associated with the cluster.)  As $\sim 22\%$ of field WDs are more
massive than 0.8\msun \citep{Liebert2005}, we would only expect to
find an average of $\sim 0.5$ massive WDs in a random field, as
opposed to the three we have identified.  The probability of finding
three high mass field WDs in our field is small ($P\sim 0.015$,
assuming Poisson statistics); it therefore seems likely that these WDs
are {\it bona fide} cluster members.

Two of the five possible cluster members (LAWDS 9 and LAWDS 20) have
masses typical of field WDs \citep[$\sim 0.56
M_{\odot}$][]{Bergeron1992}, indicating that these may be interlopers from
the field; two interlopers would be completely consistent with the 
expected number of field WDs calculated above.  It therefore
seems plausible that both lower mass candidate cluster WDs are not true
cluster members, but we cannot be certain of this with our current data.  
Proper motion measurements are needed to clarify the membership
status of these WDs.

However, as we are attempting to construct a purely empirical IFMR, it
is dangerous to throw out LAWDS 9 and LAWDS 20 simply because they do
not meet our preconceived notions of what the IFMR ought to be. We now
examine some possible explanations for how these two WDs can be
cluster members and yet have relatively low masses.

\begin{enumerate}
\item
The WDs may be members of binary systems and may have
a lowered final mass due to mass transfer onto their binary partner.
This scenario has been invoked to explain the lowest mass WDs in the
field WD population \citep[e.g.,][]{Kilic2007}.  Although these
potential NGC 1039 WDs are significantly higher mass, a similar mechanism
could still operate.  However, as the colors of these two WDs are not
abnormal for single WDs, the companion responsible for the binary
evolution either must be a very faint compact object, a very
low mass star or brown dwarf, or the companion must have been lost.

\item \citet{Kalirai2007} have shown via spectroscopic measurements
  that many WDs in the old, super metal rich open cluster
  \objectname{NGC 6791} are low-mass helium-core WDs, and have posited
  that the high metallicity of the star cluster may have led to
  enhanced mass loss during the post-main-sequence evolutionary phases
  of these stars.  A similar mechanism could be responsible for the
  apparently low masses of LAWDS 9 and LAWDS 20.
 
  However, the metallicity of NGC 1039 is consistent with solar (or
  slightly super-solar) metallicity. Furthermore, its metallicity is
  significantly lower than that of the Hyades and Praesepe clusters,
  and no abnormally low mass WDs are seen in Hyades or Praesepe.
  Enhanced mass loss therefore does not seem a likely explanation for
  the low mass of these two WDs.
\end{enumerate}

Given these points, we consider the most plausible explanation for the
comparatively low masses of LAWDS 9 and LAWDS 20 to be that these are
interlopers from the field WD population.  However, without proper
motion membership determinations, we cannot rule out that these stars
are indeed cluster members.

\section{Discussion\label{discussion}}

\subsection{Notes on Individual Objects}
Out of 44 photometrically selected WD candidates, we spectroscopically
identified 32 objects.  Twelve objects are AGNs, one is an A-type star,
and 19 are WDs.  Seventeen of these WDs are hydrogen-atmosphere DAs;
of these, five are selected to be cluster member DA WDs, and an
additional DA WD may possibly be a double-degenerate cluster member.

\emph{LAWDS 26}:
This WD exhibits  \ion{He}{1} absorption lines in its
spectrum and no hydrogen lines.  We therefore classify it as a DB WD.
The equivalent width of the \ion{He}{1} 4471 \AA~absorption line
 is $13.4$ \AA, suggesting $\teff\sim 15,000\mathrm{K}$ for $\logg=8.0$ 
\citep{Koester1980}.
 A detailed analysis of this object and its probability of cluster
membership will be presented in a future paper.

\emph{LAWDS 7}:
This object shows no sign of any atmospheric
features in the observed spectral range, so we classify it as a DC
WD.  As such, it is certainly too cool to have
been born during the lifetime of NGC 1039, so we do not consider it
a cluster member.

\emph{LAWDS S1}: Double degenerates, being a potential source of
Type Ia supernovae, are of particular interest.  At low spectral
resolutions, double DA WDs appear to be single objects with \teff\ and
\logg\ intermediate to those of the two component WDs
\citep{Bergeron1990}.  Cluster double degenerates would also appear
overluminous (to a maximum of $\approx 0.75$ mag), and would be
identified as foreground objects according to our membership criteria.
It is possible that the DA WD
LAWDS S1 may be a double degenerate.  It is $\sim 0.1$ to $\sim 0.5$
mag overluminous if at the cluster distance.  A lower WD mass could
result from close binary evolution; its initial mass, assuming that it
is a cluster member, is given in Table \ref{tab.masses}.  As the mass
of this object ($0.58 \msun$) is near the peak of the field WD mass
distribution, it seems more likely that this WD is just a solitary
field WD.

\emph{LAWDS S2}: 
We observe a very weak (EW$\approx 0.5$ \AA) \ion{He}{1} absorption
line in the spectrum of the DA LAWDS S2 at $4471$ \AA.  
No other \ion{He}{1} or \ion{He}{2} lines are
observed.  The temperature of this WD is near the red edge of the
so-called ``DB gap," the temperature region between 28,000 K and 45,000
K where very few helium-rich atmosphere WDs are known to exist.  One
explanation for the DB gap is that, in this temperature range, small
residual amounts of hydrogen in a helium-atmosphere WD form a thin
veneer obscuring the helium atmosphere; hydrogen layer masses as small as
$10^{-14}M_\odot$ are sufficient.  At 28,000 K, convection reaches the
surface layers, mixing the hydrogen back into the dominant helium
layer \citep{Fontaine1987}.  LAWDS S2 may, as indicated by its temperature 
being near the cool end of
the DB gap, be starting its transition from a DA back into a DB
spectral-type object.  Alternatively, LAWDS S2 may be a binary WD,
with DA and DB components \citep[e.g.,][]{Wesemael1994}. Additional
spectroscopy is needed to permit further analysis of this potentially
interesting object.

\subsubsection{Objects Recovered from \cite{Anthony-Twarog1982}}
\label{discussion.AT} In the early 1980s, studies of WDs in open
clusters were pioneered by several authors
\citep[e.g.,][]{Romanishin1980,Koester1981}.
\citet{Anthony-Twarog1982} included NGC 1039 in her study.  Of the
four objects identified as possible WD members of NGC 1039 by
\citet{Anthony-Twarog1982}, only one, \object[Cl* NGC 1039
A35149]{A35149} (LAWDS 30 in this work), meets our WD candidate
selection criteria.  It has been spectroscopically identified as an
AGN with a redshift $z = 0.711 \pm 0.003$ assigned from a single
emission line (\ion{Mg}{2}) identification.  The other three candidate
WD cluster members from \citet{Anthony-Twarog1982} are recovered in
our photometry but are too red to meet our WD candidate selection
criteria.  We do select several objects as candidate WDs that were
identified by \citet{Anthony-Twarog1982} but not selected in that work
as candidate WDs.  These objects are \object[Cl* NGC 1039
A43036]{A43036} (LAWDS 33, not observed spectroscopically),
\object[Cl* NGC 1039 A42085]{A42085} (LAWDS 41, an A-type star),
\object[Cl* NGC 1039 A16073]{A16073} (LAWDS 26, a DB), \object[Cl* NGC
1039 A22189]{A22189} (LAWDS 22, a DA), \object[Cl* NGC 1039
A24144]{A24144} (LAWDS 20, also \object{LB 3570}, a DA), and
\object[Cl* NGC 1039 A15118]{A15118} (LAWDS 9, also \object{LB 3567},
a DA).

\subsubsection{WDs with \ion{Ca}{2} K Absorption}

Eight of the 17 DA WDs (47\%) show significant \ion{Ca}{2} K
absorption, indicating that these are potential DAZ WDs. Measured
equivalent widths (EWs) of the \ion{Ca}{2} K absorption lines for all
WDs are given in Table \ref{tab.cak}. \citet{Zuckerman2003} find that $\sim
25\%$ of DA WDs show intrinsic \ion{Ca}{2} absorption; our rate of
incidence is significantly higher than this.  In addition, our EWs are
significantly larger than those observed in the majority of the
\citet{Zuckerman2003} sample; the higher average temperature of our WD
sample would imply enormous abundances for a large fraction of WDs,
which seems implausible.

\begin{figure*}
\plotone{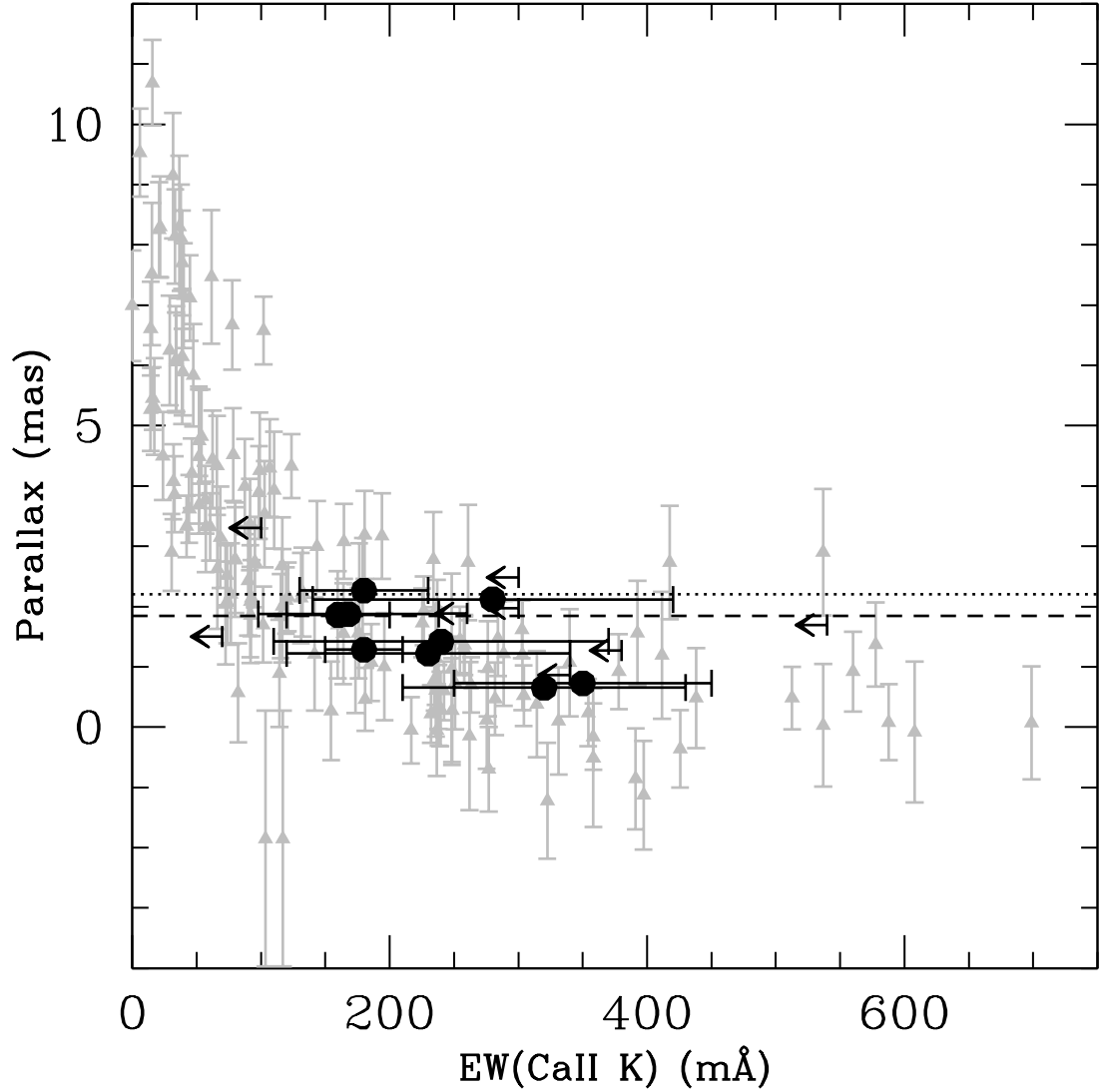}
\caption{Measurements presented in \citet[][grey triangles]{Megier2005} with our
\ion{Ca}{2} K EWs and parallaxes (computed from our calculated
$(m-M)_0$ values) overplotted.  Our vertical error bars are smaller
than the plotted point size.  $2\sigma$ EW upper limits on 
non-detections are also plotted.
Our EWs fall within the scatter of the measurements of
interstellar \ion{Ca}{2}, suggesting that \ion{Ca}{2} in the WD
spectra is interstellar in origin.  Dashed and dotted lines mark the
parallax of NGC 1039 as determined by \citet{Sarajedini2004} and 
\citet{Jones1996} respectively.}  \label{fig.megier}
\end{figure*}

There is a distinct possibility that the detected \ion{Ca}{2} is
interstellar in origin; our spectral resolution is too low to resolve
individual clouds in the interstellar medium (ISM) or to measure any
velocity differences between the Balmer lines and the \ion{Ca}{2} K
line.  
While there appears to be no significant correlation between the
\ion{Ca}{2} K EW and the derived WD distance modulus, the measured EWs
fall within the scatter of
the interstellar absorption measured in the spectra of hot stars by
\citet[see Figure~\ref{fig.megier}]{Megier2005}.  However, the
candidate DAZs do not seem to show any projected spatial correlations,
and there is no correlation between the position of these WDs and
WDs that do not exhibit \ion{Ca}{2} K absorption in the field with 
areas of enhanced diffuse H$\alpha$
emission from the maps of \citet{Finkbeiner2003}.

In short, we cannot determine conclusively whether the observed
\ion{Ca}{2} K absorption in the eight potential DAZ WD spectra is intrinsic to the
WDs, or due perhaps to patchy interstellar clouds, though the
interstellar origin seems most likely.  Higher-resolution spectroscopy
is necessary to further investigate the potential metal content of
these WDs.

\subsection{Expected Number of Cluster WDs}
The number of expected cluster WDs may be calculated by assuming an
initial mass function (IMF) for the cluster and an upper limit on the
mass of stars which form WDs; such calculations are detailed in
\citet{Williams2004}.  We calculate the expected WD population of NGC 1039
using this method.   \citet{Jones1996} combine the results
from their proper motion study of NGC 1039 with those of
\citet{Ianna1993} to construct a luminosity function for the central
$\sim 4$ parsecs of the cluster.  They find 71 cluster members with
$M_{V} < 4.0$, the completeness limit of their study.  We assume a
Salpeter IMF and an upper mass limit on WD progenitors of $8
M_{\odot}$, and find that we expect the cluster to contain 11.6 WDs in this central region.
Assuming a binary fraction of 0.5 and a random distribution of binary
mass ratios, 5.5 WDs should be detectable in the $\sim 4$ parsec
field imaged by \citet{Jones1996}; the rest will be hidden in
binary stars. 

\citet{Jones1996} assume a projected surface density profile for the
cluster of the form $\rho (r) = \rho_0 \mathrm{e}^{(r/r_0)}$, where $\rho (r)$
is in units of stars per square arcminute.  They calculate $r_0$, the
scale length for the projected surface density of cluster members, for
several different $V$ magnitude ranges.  We normalize this projected
surface density profile using the number of WDs expected in the area
imaged by \citet[calculated above]{Jones1996} and integrate it over
the area of our MOSAIC imaging.  We choose the largest value of $r_0 =
9\farcm 1$ in order to avoid underestimating the expected number of
WDs in the cluster.  The result of this integration, 7.7 WDs, is the
expected number of cluster member WDs in the field imaged in this
study.  If all six WDs observed at the cluster $(m-M)_V$ are cluster
members, this result is consistent with our findings.  If only the
three massive WDs are cluster members, then NGC 1039 may be deficient
in WDs; this deficiency has been observed in other poor open clusters
\citep[e.g.,][]{Williams2004}. We also integrate the projected surface
density profile to a radius of infinity and find that the absolute
maximum number of WDs expected is 9.9.  For smaller values of $r_0$,
the expected number of WDs decreases.

\subsection{Cluster Distance and Reddening Determination from WDs}

As noted previously, there is considerable variation in values of
the distance modulus of NGC 1039 reported in the literature.  An
alternative method of estimating cluster distances involves measuring
the distances of WDs assumed to be cluster members.  The results of
this method are independent of any assumptions we have made about the
distance of the cluster, other than the membership distance criterion
(which involved a broad selection).  

We first limit the analysis to the three massive ($M_{\rm WD} \geq
0.8M_\odot$) cluster members; the weighted average of $(m-M)_{V}$
of these objects is $8.916 \pm 0.133$.  This value matches the
\citet{Sarajedini2004} determination of $(m-M)_{V} = 8.98 \pm 0.06$
within the error bars.  If we include in the calculation the
lower mass potential cluster members, we obtain a weighted average of
$(m-M)_{V} = 8.866 \pm 0.100$, also in agreement with the
\citet{Sarajedini2004} value.
 
\citet{Sarajedini2004} use near-IR photometry from 2MASS in addition
to optical photometry in order to determine a distance modulus, as
opposed to \citet{Jones1996}, whose study is confined to $BVI$
optical photometry.  \citet{Sarajedini2004} note that when the optical
photometry alone is fitted by modern isochrones (as in
\citet{Raffauf2001}, who use \citet{Girardi2000} and \citet{Yi2001}
isochrones), a distance modulus consistent with their measurement
is obtained.  \citet{Sarajedini2004} also point out that they use a
slightly different value of $\ebv$ than in earlier studies, which
could cause them to obtain a larger value of $(m-M)_0$.

We may also use our WD model calculations to measure the extinction
toward this cluster.  The WD cooling models we have used provide us
with expected colors for cluster WDs.  These are compared with
observed colors in Table \ref{tab.reddening}.  Using only the three
high-mass cluster member WDs, the weighted average $E(B-V) = 0.111 \pm
0.025$ and $E(U-V) = 0.065 \pm 0.028$ for the cluster.  If we include
the lower mass potential cluster members, we obtain $E(B-V) = 0.099
\pm 0.021$ and $E(U-V) = 0.110 \pm 0.025$.  Both values for $E(B-V)$
are consistent with the value used by \citet{Sarajedini2004} ($E(B-V)
= 0.10 \pm 0.01$).  The values we obtain for $E(U-V)$ are lower than
$E(U-V) = 0.165$ we would expect based on the \ebv\ value and our
assumed reddening curve; however, given the uncertainty in our
$U$-band calibrations (see Table \ref{tab.coef}), we believe that this
difference is not significant.

\subsection{The Initial-Final Mass Relation\label{discussion.ifmr}}
\begin{figure*}
\plotone{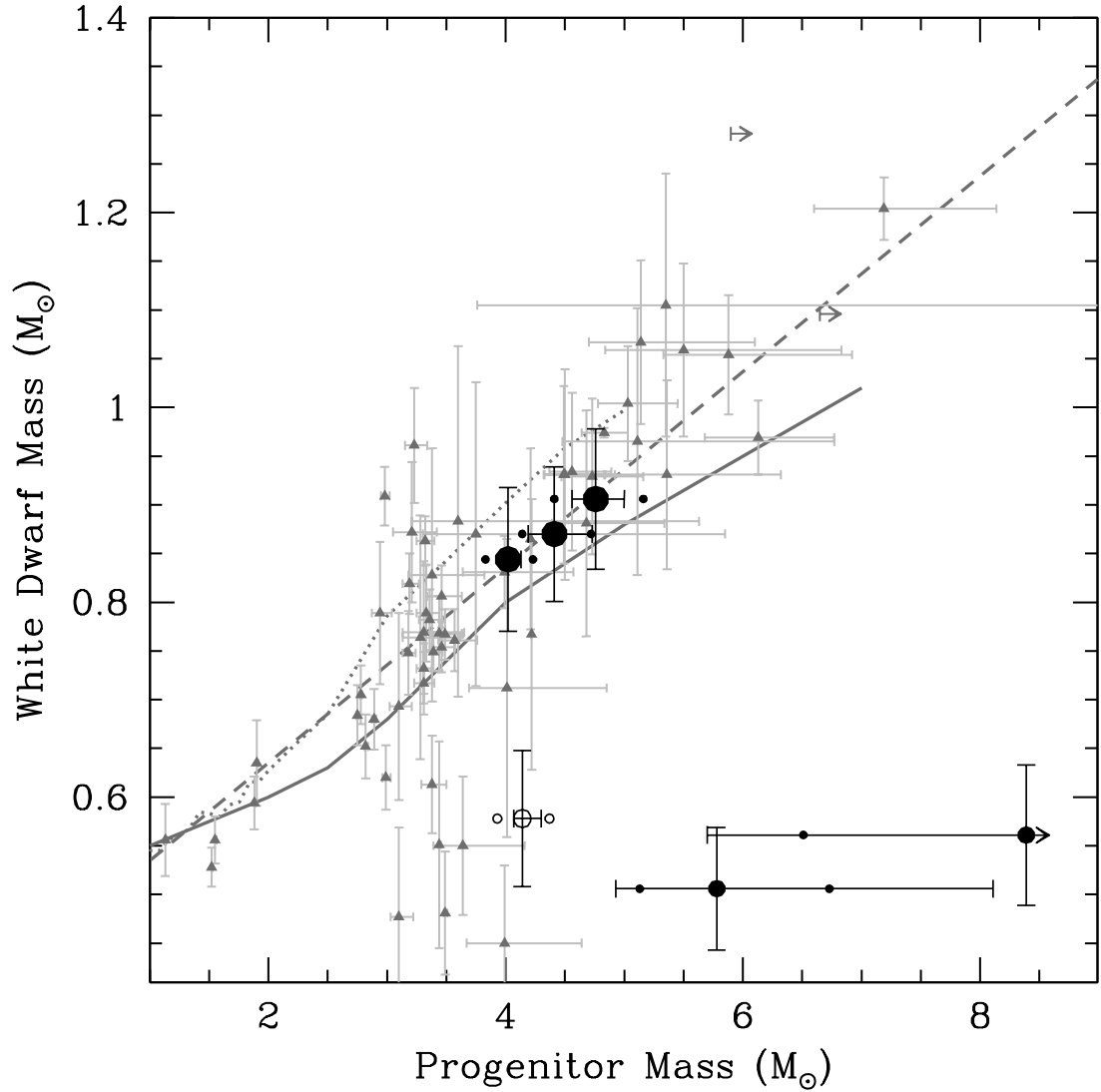}
\caption{IFMR with new results over-plotted.
Gray triangles show WDs included from the literature. High mass ($>
0.8 M_{\odot}$) NGC 1039 WDs are represented by large filled circles;
low mass NGC 1039 WDs are shown with intermediate-sized filled
circles.  The open circle represents the candidate double-degenerate binary cluster WD.
Error bars on NGC 1039 data points indicate the combined fitting
errors for each point.  Systematic changes in the location of the NGC 1039
WDs due to the uncertainty in the cluster age are shown by the small
points to the left (250 Myr) and right (200 Myr) of each
WD.  We include the linear IFMR from \citet[dashed line]{Ferrario2005}, the semi-empirical
IFMR of \citet[solid curve]{Weidemann2000}, and the theoretical, $Z=0.019$ IFMR of
\citet[dotted curve]{Marigo2001}.
\label{fig.ifmr}}
\end{figure*}

Having identified likely cluster member WDs, we now place them on the
initial-final mass relation (IFMR, Figure \ref{fig.ifmr}).  The gray
points show data from the literature, with initial and final masses
re-determined from published \teff\ and \logg\ using our adopted
evolutionary models and methodology. References to the published data
and the crucial assumptions are given in \citet{Williams2007b}.  The
high-mass WDs selected for cluster membership (LAWDS 15, 17 and S2),
represented by large black filled circles, fall directly on the trend
found by \citet{Ferrario2005} and represented by the dashed line.
LAWDS 9 and 20 (intermediate-sized filled circles) fall well below the
established IFMR, as does LAWDS S1 (the candidate double degenerate).
If cluster members, their relatively low masses compared to other WDs
with similar progenitor masses would indicate that some mechanism,
such as close binary evolution or enhanced mass loss, is necessary to
explain their location in the IFMR (see
\S\ref{discussion.contamination}).  

The three massive cluster member WDs fall almost exactly on the linear
IFMR from \citet[dashed line]{Ferrario2005}.  For illustration, we
also include the semi-empirical IFMR of \citet[solid
curve]{Weidemann2000} and the theoretical, $Z=0.019$ IFMR of
\citet[dotted curve]{Marigo2001}.  These relations bracket
the empirical data.  These points also begin to fill a region of the
diagram ($M_{\rm init}=4\msun-5\msun$) previously relatively
unpopulated.

The uncertainty in the age of the open cluster is a source of
systematic error we now consider.  As the lifetime of the progenitor
star is the cluster age minus the WD cooling age, an increase
(decrease) in the assumed cluster age results in an increase
(decrease) in the progenitor star lifetime, which corresponds to a
decrease (increase) in the progenitor star mass for each WD.  In other
words, an older assumed cluster age leads to systematically lower
progenitor mass for \emph{every} cluster WD, and vice versa.  For this
reason, we do not simply fold the cluster age uncertainty into the
individual WD error bars, where it is likely to be misinterpreted as
a random error for an individual star.  

The change in the derived IFMR
due to the uncertainty in the age of NGC 1039 is shown in Figure
\ref{fig.ifmr} as small points to the left and right of each cluster
WD, which represent the IFMR for assumed ages of 250 Myr and 200
Myr, respectively.  For this cluster, the uncertainty in the star
cluster age does not qualitatively affect the conclusion that
the NGC 1039 IFMR agrees with the existing empirical IFMR.

\begin{figure*}
\plotone{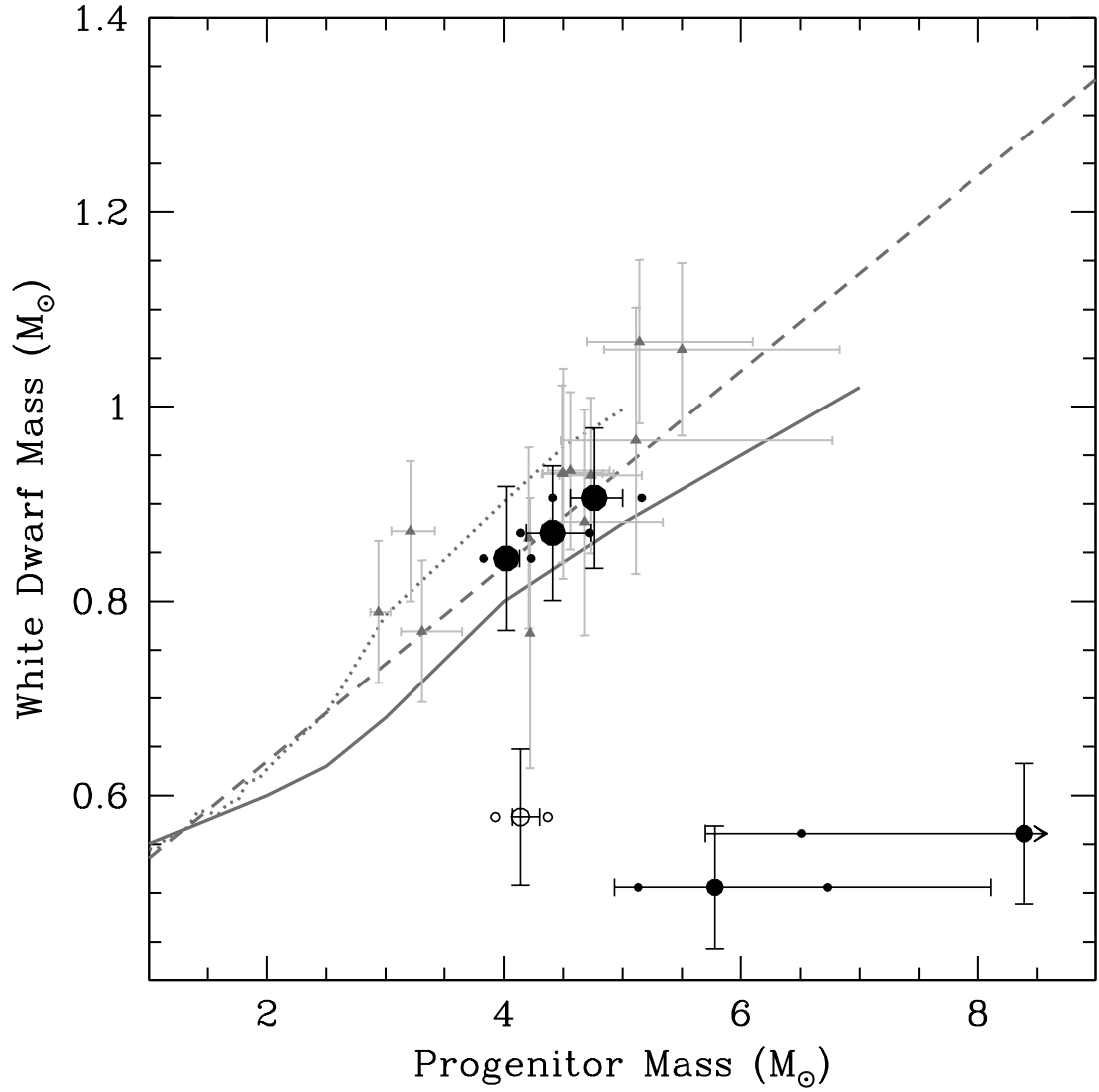}
\caption{IFMR with only points from LAWDS
plotted; symbols and lines are as in Figure \ref{fig.ifmr}.  The scatter appears
to be smaller here, perhaps indicating that some of the scatter in the
combined IFMR is due to the heterogeneity of
the data set.
\label{fig.lawds_ifmr} }
\end{figure*}

Figure \ref{fig.lawds_ifmr} shows the empirical IFMR for only those objects 
included in the LAWDS studies.  There is considerably less scatter 
around the IFMR when only these points are plotted.  This could be
indicative of the heterogeneity of the current open cluster WD sample,
which is derived from several studies using a variety of instruments
and atmospheric models.  This suggests that further study into the
inter-comparison of different samples is warranted.

\section{Conclusions}
We present photometric and spectroscopic data on the WDs in the field
of the young open cluster NGC 1039.  We use $\ubv$ imaging to
photometrically select 44 objects as WD candidates.  Of these 44, we
spectroscopically identify 32 objects.  Twelve are AGNs, one is an
A-type star, 17 are DA or DAZ WDs, one is a DB WD (LAWDS 26), and one
is a DC WD (LAWDS 7).  Five DAs are selected as possible cluster
members-- three high mass ($> 0.8 M_{\odot}$) WDs, and two low mass
WDs.  The membership of the DB WD remains in question and requires
further analysis.  The distance modulus of an additional DA is $<
0.75$ mag brighter than the distance modulus of the cluster,
making it a potential (but not highly compelling) double-degenerate
binary cluster member.

The three high-mass cluster members lie directly on the established
IFMR of \citet{Ferrario2005}, while two lower-mass candidate members
lie well below the relation.  This could be due to binary evolution or
metallicity-enhanced mass loss, though the evidence for either is
weak.  As the mass of these two latter WDs is similar to the mass at the peak of
the field WD mass distribution, and as we expect to find $\sim 2.4$
field WDs meeting our membership criteria, these lower-mass WDs may
not be true cluster members.  Proper motion measurements will be
necessary to confirm the membership of all five selected cluster
member WDs.

Based on the photometry of the five cluster member WDs, we obtain a
cluster WD distance modulus of $(m-M)_V=8.87\pm 0.10$ and a cluster
WD color excess of $\ebv=0.10\pm 0.02$.  These values are consistent
with recent measurements by
\citet[$(m-M)_V=8.98$ with $\ebv=0.10$]{Sarajedini2004}.

We note that the massive WDs at the distance modulus of the cluster
all have cooling ages shorter than the age of the cluster.  This
indicates that results from main-sequence fitting do not seriously
underestimate the cluster age.

Eight out of 17 DA WDs have detectable \ion{Ca}{2} K absorption.  The
fraction of DAs showing \ion{Ca}{2} absorption (47\%) is significantly
higher than the fraction that has been found in other studies ($\sim
25\%$).  Evidence suggests that the absorption is interstellar in origin,
although this cannot be confirmed with present data.

We use distance modulus and WD cooling time to select probable WD
cluster members, and while these criteria can provide us with a list
of strong cluster member WD candidates, a study of proper motions of
the cluster and these objects would significantly enhance our
certainty of their membership status.  We leave this for a future
work.

\acknowledgments
The authors are grateful for support for this project in the
form of the National Science Foundation AST-0307492.
KAW also acknowledges the financial support of National Science
Foundation award AST-0602288.
Any opinions, findings, and conclusions or recommendations expressed
in this material are those of the author(s) and do not necessarily
reflect the views of the National Science Foundation.

The authors wish to thank J.\ Liebert and the University of Texas
at Austin white dwarf cabal for helpful discussions during the analysis of these
results.  We also thank P.\ Bergeron, G.\ Fontaine, and M.\ Wood for
permitting use of their evolutionary models for the purposes of this
work.  The authors also thank the referee for comments that helped to
improve this paper.

The authors wish to recognize and acknowledge the very significant
cultural role and reverence that the summit of Mauna Kea has always
had within the indigenous Hawaiian community.  We are most fortunate
to have the opportunity to conduct observations from this mountain.

{\it Facilities:} \facility{Mayall(MOSAIC-1 wide-field camera)}, \facility{Nickel}, \facility{Keck:I(LRIS)}

\clearpage
\begin{deluxetable}{lccccc}
\tablecolumns{6}
\tablewidth{0pt}
\tablecaption{Transformation equation coefficients.\label{tab.coef}}
\tablehead{\colhead{Instrument} & \colhead{Band} & \colhead{Zero Point}
  & \colhead{Airmass Term} & \colhead{Color Term} & \colhead{Quadratic
    Color Term}}
\startdata
Nickel & $U$ & $A_0=21.180\pm 0.091$ & $A_1=0.419\pm 0.069$ & $A_2=-0.061\pm 0.002$ & \nodata \\
       & $B$ & $B_0=22.752\pm 0.003$ & $B_1=0.223\pm 0.008$ & $B_2=-0.083\pm 0.002$ & \nodata \\
       & $V$ & $C_0=23.014\pm 0.005$ & $C_1=0.136\pm 0.005$ & $C_2= 0.060\pm 0.001$ & \nodata \\
Mosaic & $U$ & $A_0=22.942\pm 0.016$ & \nodata              & $A_2=-0.167\pm 0.036$ & $A_3=0.085\tablenotemark{a}$ \\
       & $B$ & $B_0=25.043\pm 0.030$ & \nodata              & $B_2=-0.040\pm 0.043$ & \nodata \\
       & $V$ & $C_0=24.994\pm 0.043$ & \nodata              & $C_2= 0.013\pm 0.056$ & \nodata \\
\enddata
\tablenotetext{a}{Adopted from \citet{Williams2007b}}
\end{deluxetable}

\tabletypesize{\tiny}
\begin{deluxetable*}{lcccccccccc}
\tablecolumns{11}
\tablecaption{Photometry of candidate WDs in the field of NGC 1039.\label{tab2.phot}}
\tablewidth{0pt}
\tablehead{\colhead{LAWDS} & \colhead{R. A.} & \colhead{Decl.}& $V$ & $\sigma_V$ &  
  \colhead{$\bv$} & \colhead{$\sigma_{B-V}$} & \colhead{$U-B$} & \colhead{$\sigma_{U-B}$} & \colhead{Spec. ID} & \colhead{Comments \&}\\
  \colhead{ID} &\colhead{(J2000)} & \colhead{(J2000)} & & & & & & & & \colhead{References}}
\startdata
\objectname{NGC 1039: LAWDS 4} & 2:41:19.21 & 42:47:28.8 & 19.131 & 0.032 &  0.275 & 0.045 & $-$0.650 & 0.046 & QSO & z = 0.784, X-ray source \tablenotemark{c} \\
\objectname{NGC 1039: LAWDS 7} & 2:41:06.91 & 42:42:55.6 & 19.502 & 0.032 &  0.150 & 0.046 & $-$0.509 & 0.046 & DC & \nodata \\
\objectname{NGC 1039: LAWDS 8} & 2:41:51.59 & 42:45:28.3 & 19.587 & 0.032 &  0.079 & 0.046 & $-$0.757 & 0.046 & QSO & z = 1.262, X-ray source \tablenotemark{c} \\
\objectname{NGC 1039: LAWDS 9} & 2:40:37.77 & 42:52:29.6 & 19.628 & 0.032 &  0.119 & 0.046 & $-$0.535 & 0.046 & DA & \object{LB3567} \tablenotemark{b} \\
\objectname{NGC 1039: LAWDS 14} & 2:41:05.76 & 42:48:15.3 & 19.771 & 0.032 & $-$0.029 & 0.046 & $-$0.814 & 0.046 & DA & \object{LB3569} \tablenotemark{b} \\
\objectname{NGC 1039: LAWDS 15} & 2:40:33.73 & 42:58:16.7 & 19.806 & 0.023 & $-$0.027 & 0.032 & $-$1.067 & 0.032 & DA & \object{LB3566} \tablenotemark{b} \\
\objectname{NGC 1039: LAWDS 17} & 2:40:27.93 & 42:30:56.6 & 19.896 & 0.032 &  0.078 & 0.046 & $-$1.240 & 0.046 & DA & \object{LB3565} \tablenotemark{b} \\
\objectname{NGC 1039: LAWDS 18} & 2:40:24.77 & 42:59:33.1 & 20.106 & 0.024 &  0.105 & 0.034 & $-$0.946 & 0.033 & DA & \nodata \\
\objectname{NGC 1039: LAWDS 19} & 2:41:44.93 & 42:30:05.6 & 20.130 & 0.032 &  0.223 & 0.046 & $-$0.424 & 0.046 & DA & \nodata\\
\objectname{NGC 1039: LAWDS 20} & 2:41:09.11 & 42:43:51.1 & 20.088 & 0.032 &  0.153 & 0.046 & $-$0.445 & 0.046 & DA & \object{LB3570} \tablenotemark{b} \\
\objectname{NGC 1039: LAWDS 22} & 2:41:39.61 & 42:43:00.3 & 20.231 & 0.033 &  0.147 & 0.046 & $-$0.545 & 0.046 & DA & \object{LB3575} \tablenotemark{b} \\
\objectname{NGC 1039: LAWDS 23} & 2:40:51.68 & 42:58:33.8 & 20.204 & 0.023 &  0.151 & 0.033 & $-$0.836 & 0.033 & QSO & z = 1.949 \\
\objectname{NGC 1039: LAWDS 25} & 2:41:55.24 & 42:53:22.0 & 20.240 & 0.023 &  0.169 & 0.033 & $-$0.500 & 0.033 & DA & \object{LB3576} \tablenotemark{b} \\
\objectname{NGC 1039: LAWDS 26} & 2:42:00.22 & 42:59:48.9 & 20.355 & 0.023 &  0.005 & 0.033 & $-$0.899 & 0.033 & DB & \object[Cl* NGC 1039 A16073]{A16073} \tablenotemark{a} \\
\objectname{NGC 1039: LAWDS 30} & 2:42:55.48 & 42:59:00.8 & 20.760 & 0.023 &  0.261 & 0.034 & $-$0.495 & 0.034 & QSO & z = 0.711, \object[Cl* NGC 1039 A35149]{A35149} \tablenotemark{a} \\
\objectname{NGC 1039: LAWDS 32} & 2:42:38.40 & 42:38:46.9 & 20.776 & 0.033 &  0.171 & 0.047 & $-$0.840 & 0.047 & QSO & z = 1.509 \\
\objectname{NGC 1039: LAWDS 33} & 2:43:17.19 & 42:40:52.8 & 20.820 & 0.034 &  0.267 & 0.049 & $-$0.502 & 0.049 & \nodata  & \object[Cl* NGC 1039 A43036]{A43036} \tablenotemark{a} \\
\objectname{NGC 1039: LAWDS 34} & 2:42:59.90 & 42:38:14.3 & 20.974 & 0.034 & $-$0.073 & 0.047 & $-$1.024 & 0.046 & DA & \nodata \\
\objectname{NGC 1039: LAWDS 40} & 2:40:43.57 & 42:35:45.6 & 21.304 & 0.036 &  0.056 & 0.050 & $-$0.667 & 0.049 & DA & \nodata \\
\objectname{NGC 1039: LAWDS 41} & 2:42:33.98 & 42:37:13.3 & 15.846 & 0.032 &  0.132 & 0.045 & $-$0.655 & 0.045 & A & \object[Cl* NGC 1039 A42085]{A42085} \tablenotemark{a} \\
\objectname{NGC 1039: LAWDS 102} & 2:42:54.29 & 43:04:00.3 & 21.156 & 0.026 & $-$0.088 & 0.036 & $-$0.925 & 0.036 & DA & \nodata \\
\objectname{NGC 1039: LAWDS 103} & 2:42:58.27 & 42:53:27.8 & 21.232 & 0.026 &  0.276 & 0.040 & $-$0.533 & 0.041 & \nodata  & \nodata \\
\objectname{NGC 1039: LAWDS 104} & 2:42:54.66 & 42:40:24.3 & 21.288 & 0.036 &  0.120 & 0.050 & $-$0.590 & 0.051 & \nodata  & \nodata\\
\objectname{NGC 1039: LAWDS 105} & 2:42:29.53 & 42:38:19.4 & 21.264 & 0.035 &  0.095 & 0.049 & $-$0.506 & 0.048 & \nodata  & \nodata \\
\objectname{NGC 1039: LAWDS 107} & 2:41:42.03 & 42:38:47.7 & 21.023 & 0.034 &  0.270 & 0.048 & $-$0.517 & 0.048 & \nodata  & \nodata \\
\objectname{NGC 1039: LAWDS N3} & 2:41:11.11 & 43:13:25.3 & 18.621 & 0.032 & $-$0.217 & 0.045 & $-$1.238 & 0.045 & DA & \nodata \\
\objectname{NGC 1039: LAWDS N7} & 2:42:25.31 & 43:15:29.7 & 19.369 & 0.032 &  0.199 & 0.045 & $-$0.976 & 0.045 & QSO & z = 2.099 \\
\objectname{NGC 1039: LAWDS N8} & 2:42:46.37 & 43:12:26.4 & 19.359 & 0.033 &  0.205 & 0.046 & $-$0.698 & 0.046 & QSO & z = 1.520 \\
\objectname{NGC 1039: LAWDS N13} & 2:40:33.54 & 43:15:40.5 & 19.898 & 0.032 &  0.119 & 0.046 & $-$0.592 & 0.046 & QSO & z = 1.566 \tablenotemark{d}\\
\objectname{NGC 1039: LAWDS N15} & 2:41:33.16 & 43:18:49.3 & 20.187 & 0.032 &  0.145 & 0.046 & $-$0.862 & 0.046 & QSO & z = 1.474 \\
\objectname{NGC 1039: LAWDS N18} & 2:40:41.66 & 43:21:59.6 & 20.519 & 0.033 &  0.196 & 0.047 & $-$0.656 & 0.047 & QSO & z = 1.824 \\
\objectname{NGC 1039: LAWDS N19} & 2:41:16.76 & 43:12:11.3 & 20.524 & 0.033 &  0.173 & 0.047 & $-$0.458 & 0.048 & QSO & z = 1.686 \\
\objectname{NGC 1039: LAWDS N20} & 2:43:29.02 & 43:05:10.4 & 21.054 & 0.034 &  0.072 & 0.049 & $-$0.664 & 0.049 & \nodata  & \nodata \\
\objectname{NGC 1039: LAWDS N21} & 2:42:21.98 & 43:19:25.1 & 21.366 & 0.036 &  0.135 & 0.053 & $-$0.972 & 0.052 & \nodata  & \nodata \\
\objectname{NGC 1039: LAWDS N22} & 2:41:59.84 & 43:22:56.4 & 21.281 & 0.035 &  0.031 & 0.052 & $-$1.207 & 0.051 & \nodata  & \nodata \\
\objectname{NGC 1039: LAWDS S1} & 2:41:17.12 & 42:25:46.8 & 18.965 & 0.032 & $-$0.044 & 0.046 & $-$0.875 & 0.046 & DA & \nodata \\
\objectname{NGC 1039: LAWDS S2} & 2:41:05.05 & 42:15:59.0 & 19.330 & 0.033 & $-$0.145 & 0.046 & $-$1.183 & 0.046 & DA & \nodata \\
\objectname{NGC 1039: LAWDS S3} & 2:40:59.08 & 42:15:13.5 & 19.534 & 0.033 &  0.183 & 0.047 & $-$0.546 & 0.047 & DA & \nodata \\
\objectname{NGC 1039: LAWDS S4} & 2:41:47.60 & 42:17:16.5 & 20.181 & 0.034 &  0.166 & 0.050 & $-$0.703 & 0.049 & QSO & z = 1.445 \\
\objectname{NGC 1039: LAWDS S5} & 2:41:33.01 & 42:03:47.3 & 20.951 & 0.042 &  0.028 & 0.063 & $-$0.463 & 0.060 & DA & \nodata \\
\objectname{NGC 1039: LAWDS S6} & 2:43:24.85 & 42:09:34.5 & 19.914 & 0.034 &  0.238 & 0.049 & $-$0.550 & 0.048 & \nodata  & \nodata \\
\objectname{NGC 1039: LAWDS S7} & 2:43:05.14 & 42:06:24.1 & 20.495 & 0.036 &  0.211 & 0.053 & $-$0.785 & 0.051 & \nodata  & \nodata \\
\objectname{NGC 1039: LAWDS S10} & 2:41:09.14 & 42:07:48.6 & 21.327 & 0.059 & $-$0.100 & 0.086 & $-$0.608 & 0.073 & \nodata  & \nodata\\
\objectname{NGC 1039: LAWDS S11} & 2:40:29.04 & 42:25:51.4 & 20.858 & 0.041 &  0.262 & 0.069 & $-$0.580 & 0.068 & \nodata  & \nodata \\
\enddata
\tablecomments{Units of right ascension are hours, minutes and seconds, and units of declination are degrees, arcminutes, and arcseconds.  Identifications of ``A'' indicate non-WD spectra with Balmer-series lines.}
\tablenotetext{a}{Object from \citet{Anthony-Twarog1982}}
\tablenotetext{b}{Object from \citet{Luyten1961}}
\tablenotetext{c}{X-ray source from \citet{Simon2000}} 
\tablenotetext{d}{Uncertain redshift measurement}
\end{deluxetable*}
\clearpage

\clearpage
\begin{landscape}
\tabletypesize{\footnotesize}
\begin{deluxetable*}{lccccccccccccccc}
\tablecolumns{16}
\tablecaption{Spectral fits for DA WDs in the field of NGC 1039.\label{tab.spec}}
\tablewidth{0pt}
\tablehead{\colhead{LAWDS} & \colhead{S/N \tablenotemark{a}} & \colhead{\teff} & \colhead{$\sigma_{\teff, {\rm int}}$} & \colhead{$\sigma_{\teff, {\rm tot}}$} & \colhead{\logg} & \colhead{$\sigma_{\logg, {\rm int}}$} & \colhead{$\sigma_{\logg, {\rm tot}}$} &
   \colhead{$M_{\mathrm WD}$} & \colhead{$\sigma_{M_{\mathrm WD}}$} & \colhead{$\tau_{\mathrm WD}$} & \colhead{$\sigma_{\tau_{\mathrm WD}}$} &  \colhead{$M_V$} & \colhead{$\sigma_{M_V}$} & \colhead{$(m-M)_V$} & \colhead{$\sigma_{(m-M)_V}$}\\
   \colhead{ID} & \colhead{} & \colhead{K} & \colhead{K} & \colhead{K}
   & & & & \colhead{$M_{\odot}$} & \colhead{$M_{\odot}$} & \colhead{log(yr)} & \colhead{log(yr)} & & & & } 

\startdata
LAWDS 9 & 115 & 15300 &  70 & 1100 & 7.82 & 0.01 & 0.12 & 0.506 & 0.063 & 8.141 & 0.132 & 10.949 & 0.210 &  8.679 & 0.212\\
LAWDS 14 & 118 & 21100 &  50 & 1100 & 7.80 & 0.02 & 0.12 & 0.514 & 0.059 & 7.537 & 0.130 & 10.348 & 0.203 &  9.423 & 0.206\\
LAWDS 15 & 175 & 25900 &  50 & 1100 & 8.38 & 0.01 & 0.12 & 0.870 & 0.074 & 7.791 & 0.181 & 10.861 & 0.231 &  8.945 & 0.232\\
LAWDS 17 & 189 & 24700 &  70 & 1100 & 8.44 & 0.01 & 0.12 & 0.906 & 0.072 & 7.949 & 0.158 & 11.058 & 0.231 &  8.838 & 0.233\\
LAWDS 18 & 219 & 19400 &  50 & 1100 & 7.84 & 0.01 & 0.12 & 0.529 & 0.065 & 7.734 & 0.147 & 10.559 & 0.207 &  9.547 & 0.208\\
LAWDS 19 & 118 & 11300 &  50 & 1100 & 8.02 & 0.01 & 0.12 & 0.617 & 0.079 & 8.652 & 0.132 & 11.796 & 0.313 &  8.334 & 0.315\\
LAWDS 20 & 110 & 14700 &  50 & 1100 & 7.92 & 0.01 & 0.12 & 0.561 & 0.072 & 8.269 & 0.135 & 11.157 & 0.219 &  8.931 & 0.221\\
LAWDS 22 &  92 & 18800 &  50 & 1100 & 7.72 & 0.01 & 0.12 & 0.465 & 0.055 & 7.711 & 0.131 & 10.444 & 0.211 &  9.787 & 0.214\\
LAWDS 25 &  79 & 15400 &  50 & 1100 & 7.92 & 0.01 & 0.12 & 0.562 & 0.072 & 8.201 & 0.129 & 11.076 & 0.204 &  9.164 & 0.205\\
LAWDS 34 &  70 & 27700 & 160 & 1110 & 7.76 & 0.04 & 0.13 & 0.512 & 0.060 & 7.068 & 0.064 &  9.747 & 0.228 & 11.227 & 0.231\\
LAWDS 40 &  37 & 19000 & 160 & 1110 & 7.90 & 0.04 & 0.13 & 0.565 & 0.078 & 7.838 & 0.164 & 10.683 & 0.229 & 10.621 & 0.232\\
LAWDS 102 &  71 & 24200 & 100 & 1100 & 7.84 & 0.02 & 0.12 & 0.541 & 0.065 & 7.300 & 0.112 & 10.151 & 0.212 & 11.005 & 0.214\\
LAWDS N3 & 239 & 44000 & 130 & 1110 & 7.74 & 0.01 & 0.12 & 0.558 & 0.049 & 6.504 & 0.050 &  8.856 & 0.210 &  9.765 & 0.212\\
LAWDS S1 & 170 & 22200 & 130 & 1110 & 7.92 & 0.02 & 0.12 & 0.578 & 0.070 & 7.529 & 0.156 & 10.432 & 0.208 &  8.533 & 0.210\\
LAWDS S2 & 201 & 31200 & 100 & 1100 & 8.32 & 0.03 & 0.12 & 0.844 & 0.074 & 7.287 & 0.215 & 10.368 & 0.223 &  8.962 & 0.225\\
LAWDS S3 & 170 & 14700 & 100 & 1100 & 8.36 & 0.03 & 0.12 & 0.842 & 0.073 & 8.570 & 0.123 & 11.818 & 0.231 &  7.716 & 0.233\\
LAWDS S5 &  67 & 15900 & 200 & 1120 & 7.96 & 0.03 & 0.12 & 0.587 & 0.072 & 8.182 & 0.132 & 11.077 & 0.204 &  9.874 & 0.208\\

\enddata
\tablenotetext{a}{Average S/N per resolution element in the pseudocontinuum surrounding H$\delta$}
\end{deluxetable*}
\clearpage
\end{landscape}

\tabletypesize{\footnotesize}
\begin{deluxetable*}{lccccc}
\tablecolumns{6}
\tablecaption{Progenitor masses for NGC 1039 WDs.\label{tab.masses}}
\tablewidth{0pt}
\tablehead{\colhead{LAWDS} & \colhead{$M_{\mathrm WD}$ \tablenotemark{a}} & \colhead{$\sigma_{M_{\mathrm WD}}$ \tablenotemark{a}} & \multicolumn{3}{c}{$M_{init}\,(M_\odot)$}\\ \cline{4-6}
        \colhead{ID} & \colhead{$M_{\odot}$} & \colhead{$M_{\odot}$} & \colhead{$\log\tau_{cl}=8.30$} & \colhead{$\log\tau_{cl}=8.35$} & \colhead{$\log\tau_{cl}=8.40$}}
        
\startdata
LAWDS 9 & 0.506 & 0.063 &  6.73 & $ 5.78^{+2.33}_{-0.85} $ & 5.13 \\
LAWDS 15 & 0.870 & 0.074 &  4.72 & $ 4.41^{+0.32}_{-0.22} $ & 4.14 \\
LAWDS 17 & 0.906 & 0.072 &  5.16 & $ 4.76^{+0.24}_{-0.20} $ & 4.41 \\
LAWDS 20 & 0.561 & 0.072 & 18.38 & $ 8.39^{+9.99}_{-2.69} $ & 6.51 \\
LAWDS S1 \tablenotemark{b} & 0.578 & 0.070 &  4.37 & $ 4.14^{+0.16}_{-0.07} $ & 3.93 \\
LAWDS S2 & 0.844 & 0.074 &  4.23 & $ 4.02^{+0.11}_{-0.06} $ & 3.83 \\

\enddata
\tablenotetext{a}{From Table~\ref{tab.spec}}
\tablenotetext{b}{Potential binary member}
\end{deluxetable*}

\tabletypesize{\footnotesize}
\begin{deluxetable*}{lccc}
\tablecolumns{3}
\tablecaption{\ion{Ca}{2} K EW measurements for WDs in the field of NGC 1039.\label{tab.cak}}
\tablewidth{0pt}
\tablehead{\colhead{LAWDS} & \colhead{\ion{Ca}{2} K EW\tablenotemark{a}} & \colhead{dEW}\\
        \colhead{ID} & \colhead{m\AA} &  \colhead{m\AA} }
\startdata
LAWDS 7 &       $ < 147$ &  \nodata  \\
LAWDS 9 &       280      & 140\\
LAWDS 14 &      $ < 70$  &   \nodata \\
LAWDS 15 &       168     & 70 \\
LAWDS 17 &      $ < 300$ &  \nodata  \\  
LAWDS 18 &       240     & 130\\
LAWDS 19 &      $ < 300$ &  \nodata  \\
LAWDS 20 &      $ < 260$ &  \nodata  \\        
LAWDS 22 &      $ < 380$ &  \nodata  \\
LAWDS 25 &      $ < 540$ &  \nodata  \\
LAWDS 26 &       230     & 100\\
LAWDS 34 &       320     & 110\\
LAWDS 40 &      $ < 340$ &  \nodata  \\
LAWDS 102 &      350     & 100\\
LAWDS N3 &       180     & 30 \\
LAWDS S1 &       180     & 50 \\
LAWDS S2 &       160     & 40 \\
LAWDS S3 &      $ < 100$ &  \nodata \\
LAWDS S5 &       230     & 110\\
\enddata
\tablenotetext{a}{Upper limits on EWs are at the $2\sigma$ level.} 
\end{deluxetable*}

\tabletypesize{\footnotesize}
\begin{deluxetable*}{lccccccc}
\tablecolumns{8}
\tablecaption{Comparison of WD model colors with observed colors.\label{tab.reddening}}
\tablewidth{0pt}
\tablehead{\colhead{LAWDS} & \colhead{$(m-M)_{V}$} & \colhead{$(B-V)$ \tablenotemark{c}} & \colhead{$(B-V)_{0}$ \tablenotemark{d}} & \colhead{$E(B-V)$} & \colhead{$(U-V)$ \tablenotemark{c}} & \colhead{$(U-V)_{0}$ \tablenotemark{d}} & \colhead{$E(U-V)$}}
\startdata
LAWDS 9 \tablenotemark{a} &  8.679 &  0.119 &  0.054 &  0.065 & -0.416 & -0.648 &  0.232 \\
LAWDS 15 &  8.945 & -0.027 & -0.128 &  0.101 & -1.094 & -1.210 &  0.116 \\
LAWDS 17 &  8.838 &  0.078 & -0.108 &  0.186 & -1.162 & -1.167 &  0.005 \\
LAWDS 20 \tablenotemark{a} &  8.931 &  0.153 &  0.084 &  0.069 & -0.292 & -0.603 &  0.311 \\
LAWDS S1 \tablenotemark{b} &  8.533 & -0.044 & -0.092 &  0.048 & -0.919 & -1.061 &  0.142 \\
LAWDS S2 &  8.962 & -0.145 & -0.201 &  0.056 & -1.328 & -1.374 &  0.046 \\
\enddata
\tablenotetext{a}{Potential low mass member}
\tablenotetext{b}{Potential binary member}
\tablenotetext{c}{Observed color, from Table \ref{tab2.phot}}
\tablenotetext{d}{Color determined from WD cooling models.}
\end{deluxetable*}
\clearpage

\end{document}